\documentclass[authoryear]{elsarticle}

\usepackage{amsmath}
\usepackage{graphicx,psfrag,epsf}
\usepackage{enumerate}
\usepackage{hyperref}

\usepackage{subcaption}
\usepackage{amssymb}
\usepackage{algorithm}
\usepackage{algorithmicx}
\usepackage{algcompatible}
\usepackage{algpseudocode}
\usepackage[table]{xcolor}

\newtheorem{assumption}{Assumption}[section]
\newtheorem{property}{Property}[section]

\newcounter{probnum}
\allowdisplaybreaks

\definecolor{gray}{rgb}{0.7,0.7,0.7}
\definecolor{black}{rgb}{0,0,0}
\newcommand{\ingray}[1]{\color{gray}#1\color{black}}

\DeclareMathOperator*{\argmax}{arg\,max~}

\newcommand{\qu}[1]{``{#1}''}
\newcommand{\bv}[1]{\boldsymbol{#1}}

\newcommand{\sigsq}{\sigma^2}

\newcommand{\sigsqzsq}{\parens{\sigma_z^2}^2}

\newcommand{\bSigmaw}{\bv{\Sigma}_{W}}
\newcommand{\bSigmawhat}{\hat{\bv{\Sigma}}_{W}}

\newcommand{\thetahat}{\hat{\theta}}
\newcommand{\betaThat}{\hat{\beta}_T}
\newcommand{\betaThatDiffMeans}{\hat{\beta}_T^{\text{DM}}}
\newcommand{\betaThatDiffMeanssq}{\hat{\beta}_T^{{\text{DM}^2}}}
\newcommand{\betaThatLinRegr}{\hat{\beta}_T^{\text{LR}}}

\newcommand{\Ybar}{\bar{Y}}
\newcommand{\YbarT}{\Ybar_T}
\newcommand{\YbarC}{\Ybar_C}
\newcommand{\xbar}{\bar{x}}

\newcommand{\zbar}{\bar{z}}

\newcommand{\allocspace}{\mathbb{W}}
\newcommand{\allocspacebase}{\allocspace_{\text{base}}}
\newcommand{\iid}{~{\buildrel iid \over \sim}~}

\newcommand{\half}{\frac{1}{2}}

\newcommand{\A}{\bv{A}}

\newcommand{\B}{\bv{B}}
\newcommand{\D}{\bv{D}}

\newcommand{\G}{\bv{G}}
\newcommand{\K}{\bv{K}}

\newcommand{\R}{\bv{R}}
\renewcommand{\r}{\bv{r}}

\newcommand{\Z}{\bv{Z}}
\newcommand{\X}{\bv{X}}
\newcommand{\Xtilde}{\tilde{\X}}

\newcommand{\I}{\bv{I}}

\newcommand{\Y}{\bv{Y}}

\renewcommand{\P}{\bv{P}}

\newcommand{\Xt}{\bv{X}^T}
\newcommand{\Xtildet}{\Xtilde^T}

\newcommand{\XtXinv}{\parens{\Xt\X}^{-1}}
\newcommand{\XtXminussqrt}{\parens{\Xt\X}^{-\half}}
\newcommand{\XtXinvXt}{\XtXinv\Xt}
\newcommand{\XXtXinvXt}{\X\XtXinvXt}

\newcommand{\XtXinvtilde}{\parens{\Xtildet\Xtilde}^{-1}}
\newcommand{\XtXinvXttilde}{\XtXinvtilde\Xtildet}

\newcommand{\BXprime}{\B_{\X}'}
\newcommand{\BXprimet}{\BXprime^\top}

\newcommand{\x}{\bv{x}}

\newcommand{\w}{\bv{w}}
\newcommand{\W}{\bv{W}}
\newcommand{\onevec}{\bv{1}}
\newcommand{\zerovec}{\bv{0}}

\newcommand{\y}{\bv{y}}

\renewcommand{\v}{\bv{v}}

\newcommand{\z}{\bv{z}}

\newcommand{\bbeta}{\bv{\beta}}

\newcommand{\twovec}[2]{\bracks{\begin{array}{c} #1 \\ #2 \end{array}}}

\newcommand{\twobytwomat}[4]{\bracks{\begin{array}{cc} #1 & #2 \\ #3 & #4 \end{array}}}

\newcommand{\reals}{\mathbb{R}}

\newcommand{\beqn}{\vspace{-0.25cm}\begin{eqnarray*}}
\newcommand{\eeqn}{\end{eqnarray*}}
\newcommand{\bneqn}{\vspace{-0.25cm}\begin{eqnarray}}
\newcommand{\eneqn}{\end{eqnarray}}
\newcommand{\benum}{\begin{enumerate}}
\newcommand{\eenum}{\end{enumerate}}

\newcommand{\parens}[1]{\left(#1\right)}
\newcommand{\squared}[1]{\parens{#1}^2}
\newcommand{\angbrace}[1]{\left<#1\right>}
\newcommand{\tothepow}[2]{\parens{#1}^{#2}}
\newcommand{\prob}[1]{\mathbb{P}\parens{#1}}

\newcommand{\cprob}[2]{\prob{#1~|~#2}}

\newcommand{\sumonen}[2]{\sum_{#1=1}^n #2}

\newcommand{\bracks}[1]{\left[#1\right]}
\newcommand{\braces}[1]{\left\{#1\right\}}

\newcommand{\abss}[1]{\left|#1\right|}
\newcommand{\norm}[1]{\left|\left|#1\right|\right|}
\newcommand{\normsq}[1]{\norm{#1}^2}
\newcommand{\normfourth}[1]{\norm{#1}^4}
\newcommand{\frob}[1]{\norm{#1}_F}
\newcommand{\frobsq}[1]{\normsq{#1}_F}

\newcommand{\tr}[1]{\text{tr}\bracks{#1}}

\newcommand{\expe}[1]{\mathbb{E}\bracks{#1}}
\newcommand{\cexpenostr}[2]{\mathbb{E}[#1\,|\,#2]}
\newcommand{\cexpe}[2]{\expe{#1\,|\,#2}}
\newcommand{\cexpesub}[3]{\expesub{#1}{#2\,|\,#3}}
\newcommand{\cexpesubnostr}[3]{\mathbb{E}_{\,#1}[#2\,|\,#3]}

\newcommand{\cvar}[2]{\var{#1\,|\,#2}}

\newcommand{\cvarsubnostr}[3]{\mathbb{V}\text{ar}_{#1}[#2\,|\,#3]}

\newcommand{\expesub}[2]{\mathbb{E}_{\,#1}\bracks{#2}}
\newcommand{\expesubnostr}[2]{\mathbb{E}_{\,#1}[#2]}

\newcommand{\var}[1]{\mathbb{V}\text{ar}\bracks{#1}}
\newcommand{\varsub}[2]{\mathbb{V}\text{ar}_{#1}\bracks{#2}}

\newcommand{\varnostrsub}[2]{\mathbb{V}\text{ar}_{#1}[#2]}

\newcommand{\cmsesub}[3]{\mathbb{M}\text{SE}_{#1}\bracks{#2\,|\,#3}}
\newcommand{\cmsesubnostr}[3]{\mathbb{M}\text{SE}_{#1}[#2\,|\,#3]}

\newcommand{\se}[1]{\mathbb{S}\text{E}\bracks{#1}}
\newcommand{\sesub}[2]{\mathbb{S}\text{E}_{#1}\bracks{#2}}

\newcommand{\expnostr}[1]{\mathrm{exp}(#1)}

\newcommand{\oneover}[1]{\frac{1}{#1}}

\newcommand{\overn}[1]{\frac{#1}{n}}

\newcommand{\multnormnot}[3]{\mathcal{N}_{#1}\parens{#2,\,#3}}

\newcommand{\chisq}[1]{\chi^2_{#1}}

\newcommand{\betaT}{\beta_T}

\newcommand{\Oof}[1]{O\parens{#1}}

\newcommand{\imbal}{{\cal B}}

\begin{document}

\begin{frontmatter}

\title{Optimal Rerandomization Designs via a Criterion that Provides Insurance Against Failed Experiments}
\author{Adam Kapelner\corref{mycorrespondingauthor}}
\cortext[mycorrespondingauthor]{Corresponding author}
\address{Department of Mathematics, Kiely Hall Room 604, 65-30 Kissena Boulevard, Queens, NY, 11367, USA}
\ead{kapelner@qc.cuny.edu}

\author{Abba M. Krieger}
\address{Department of Statistics, Huntsman Hall Room 442, 3730 Walnut Street, Philadelphia, PA, 19104, USA}

\author{Michael Sklar}
\address{Department of Statistics, Sequoia Hall Room 105, 390 Serra Mall, Stanford, CA, 94305, USA}

\author{David Azriel}
\address{Faculty of Industrial Engineering and Management, Bloomfield Building Room 301, Technion City, Haifa 32000 Israel}



\begin{abstract}
We present an optimized rerandomization design procedure for a non-sequential treatment-control experiment. Randomized experiments are the gold standard for finding causal effects in nature. But sometimes random assignments result in unequal partitions of the treatment and control group visibly seen as imbalance in observed covariates. There can additionally be imbalance on unobserved covariates. Imbalance in either observed or unobserved covariates increases treatment effect estimator error inflating the width of confidence regions and reducing experimental power. \qu{Rerandomization} is a strategy that omits poor imbalance assignments by limiting imbalance in the observed covariates to a prespecified threshold. However, limiting this threshold too much can increase the risk of contracting error from unobserved covariates. We introduce a criterion that combines observed imbalance while factoring in the risk of inadvertently imbalancing unobserved covariates. We then use this criterion to locate the optimal rerandomization threshold based on the practitioner's level of desired insurance against high estimator error. We demonstrate the gains of our designs in simulation and in a dataset from a large randomized experiment in education. We provide an open source \texttt{R} package available on \texttt{CRAN} named \texttt{OptimalRerandExpDesigns} which generates designs according to our algorithm.
\end{abstract}


\end{frontmatter}


\section{Background}\label{sec:intro}

We consider a classic problem: a treatment-control experiment with  $n$ \emph{subjects} (\emph{individuals}, \emph{participants} or \emph{units}) and one clearly defined \emph{outcome} of interest (also called the \emph{response} or \emph{endpoint}) for the $n$ subjects denoted $\y = \bracks{y_1, \ldots, y_n}^\top$. We scope our discussion to the response being continuous and uncensored (with incidence and survival response left to further research). 

Each subject is assigned to the treatment group and a control group denoted $T$ and $C$ and referred to as the two \emph{arms}. Here we consider the settings where all subjects along with their observed subject-specific \emph{covariates} (\emph{measurements} or \emph{characteristics}) denoted $\x$ known beforehand and considered fixed. This non-sequential setting was studied by \citet{Fisher1925} when assigning treatments to agricultural plots and this setting is still of great importance today. In fact, it occurs in clinical trials as \qu{many phase I studies use `banks' of healthy volunteers ... [and] ... in most cluster randomised trials, the clusters are identified before treatment is started} \citep[page 1440]{Senn2013}.

Synonymously referred to as a \emph{randomization}, an \emph{allocation} or an \emph{assignment} is a vector $\w = \bracks{w_1, \ldots, w_n}^\top$ whose entries indicate whether the subject received $T$ (coded as +1) or $C$ (coded as -1) and thus this vector must be an element in $\braces{-1,+1}^n$.

The goals of such experiments are to estimate and test a population average treatment effect (PATE) denoted $2\betaT$  (where the multiplicative factor of 2 would not be present if $\w$ was coded with 0 and 1).

After the sample is provided, the only degree of control the experimenter has is to choose the entries of $\w$. The process that results in the choices of such allocations of the $n$ subjects to the two arms is termed synonymously as an \emph{experimental design} (or just \emph{design}), a \emph{strategy}, an \emph{algorithm}, a \emph{method} or a \emph{procedure}. The design essentially is a generalized multivariate Bernoulli random variable denoted $W$ with mass function denoted $\prob{\w}$, variance denoted $\bSigmaw$ and support denoted $\allocspace$ and synonymously termed the \emph{allocation space}. If all realizations satisfy $\w^\top \onevec_n = 0$, i.e. all of the assignments feature an equal number of subjects assigned to the treatment group and the control group, $n/2$, we term $W$ a \emph{forced balance procedure} \citep[Chapter 3.3]{Rosenberger2016}.

The question then becomes: what is the optimal design strategy? Regardless of how optimality is defined on whichever criterion, this is an impossible question to answer. The multivariate Bernoulli random variable has $2^n - 1$ parameters \citep[Section 2.3]{Teugels1990}. Finding the optimal design is tantamount to solving for exponentially many parameters in $n$ using only the $n$ observations, a hopelessly unidentifiable task. Thus all \qu{optimal} designs will fundamentally be heuristics.

The question then becomes: which heuristic design should we employ? This has been hotly debated for over 100 years. There are loosely two main camps: those that optimize assignments and those that randomize assignments. The latter first advocated by \citet{Fisher1925} has prevailed, certainly in the context of causal inference and clinical trials. However, the design being \qu{best} is dependent on many choices: 

\begin{enumerate}
\item[(C1)] the choice of estimator for $\betaT$, 
\item[(C2)] the source of the randomness in the response-generating process $\Y$, 
\item[(C3)] the form of the response model as a function of the observed and unobserved covariates (and embedded within is the importance of the observed covariates relative to the unobserved) and
\item[(C4)] a definition of optimality and a loss function that gauges this optimality.
\end{enumerate}

Further, drawing valid inference will be different depending on these choices, a point that we address in Section~\ref{subsec:inference}, but here we will continue this example with the focus on estimation. We now briefly describe our (C1-C3). (C4) will be developed iteratively in the next section of the paper.

For (C1), we examine two choices of estimators for $\betaT$. First, the classic difference in means estimator $\betaThatDiffMeans := (\YbarT - \YbarC)/2$ where $\YbarT$ is the random variable of the average response in the treatment group and $\YbarC$ is the random variable of the average response in the treatment group. If the model were to be unknown, $\betaThatDiffMeans$ would be the uniformly minimum variance unbiased estimator for the PATE \citep[Example 4.7]{Lehmann1998}. Second, the covariate-adjusted \emph{regression estimator} or \emph{OLS estimator}, $\betaThatLinRegr$ defined as the slope coefficient of $\w$ in an ordinary least squares fit of $\y$ using $\x$ and $\w$. If $\y$ is known to be linear in $\x$, then $\betaThatLinRegr$ is known to be the best linear unbiased estimator for the PATE.\footnote{The employment of  $\betaThatLinRegr$ in the realistic setting where the linear model assumption is untestable (and most likely incorrect) results in bias. This bias is a subject of a large debate \citep{Freedman2008,Berk2010,Lin2013} but is known to disappear rapidly in sample size.} 

In (C2), the source of randomness in $\Y$ is typically assumed to either be the \emph{population model} \citep[Chapter 6.2]{Rosenberger2016} or the \emph{randomization model} \citep[Chapter 6.3]{Rosenberger2016}. To explain the difference between these two, assume for (C3) the following linear response model which we me generalize later:

\bneqn\label{eq:simple_model}
\y = \betaT \w + \beta_x \x + \z
\eneqn

To make the expressions even simpler, we assume $\x$ is centered and scaled so its entries have average zero and standard deviation one. This normalization allows us to omit an intercept term in this simple response model without consequences to this discussion. 

In the \emph{population model}, subjects in the treatment group are sampled at random from an infinitely-sized superpopulation and thus $\z$ is a realization from some data generating process $\Z$ while $\w$ is fixed. In the \emph{randomization model}, it is the opposite: the source of the randomness is in the treatment assignments $\w$ which is drawn from $\W$ while $\z$ is fixed. 

These two choices for (C2) represent very different worldviews with a rich history of debate \citep[see for instance][pages 125-127]{Reichardt1999}. We touch on some of these issues by elaborating on the vastly different properties of our two estimators for $\betaT$ in \ref{app:population_model} and \ref{app:randomization_model}. 

\subsection{Assumption of the Randomization Model}\label{subsec:rand_mod_assumption}

In this work, we employ the randomization model perspective following \citet{Rosenberger2016, Freedman2008, Lin2013} and others mostly because the population model requires the subjects to truly be sampled at random from a superpopulation. However, in most experiments, subjects are recruited from nonrandom sources in nonrandom locations. Experimental settings are frequently selected because of expertise, an ability to recruit subjects, and their budgetary requirements \citep[page 99]{Rosenberger2016}. In the context of clinical trials, \citet[page 296]{Lachin1988} states rather harshly that, \qu{the invocation of a population model for the analysis of a clinical trial becomes a matter of faith that is based upon assumptions that are inherently untestable}.\footnote{However, assuming the randomization model complicates the claim of the generalizability of the study's results. For reasons why these complications may be ignored, see the discussion in \citep[Chapter 6.3]{Rosenberger2016}.}

\subsection{Restricted Designs}\label{subsec:restricted_designs}

We now return to our initial question of \qu{which heuristic design should we employ?} and develop more terminology. If we were to assign each individual to treatment by an independent fair coin flip we will term this the \emph{the Bernoulli Trial} \citep[Chapter 4.2]{Imbens2015}. In the Bernoulli Trial, all assignments are possible i.e. $\allocspace = \braces{-1, 1}^n$ and all assignments (the $n$ elements of $W$) are independent. Thus the variance of the design $W$ is simply $\bSigmaw = \I_n$ where the latter notation indicates the $n \times n$ identity matrix. 

Any other design besides for this Bernoulli Trial is conventionally termed a \emph{restricted randomization} because the permissible allocations are restricted to a subset of all possible allocations. A very weak restriction is that of forced balance. If each forced balance allocation is equally likely, we term this as the \emph{balanced completely randomized design} (BCRD) as in \citet{Wu1981}. Here, there is a slight, but vanishing negative covariance between the assignments and thus the off-diagonal elements of $\bSigmaw$ are all $-\tothepow{n-1}{-1}$. 

However, there is a large problem with employing Bernoulli Trial or BCRD assignments described at the outset: under some \qu{unlucky} assignments $\w$ there are unluckily large differences in the distribution of observed covariates between the two groups creating error in the estimation. Large differences are destructive for either choice of estimator (C1) and either choice of source of randomness model (C2) as shown explicitly in \ref{app:unlucky_assignments}. The solution is to eliminate these unlucky assignments from $\allocspace$, the main reason restricted randomization has been employed for the past 100 years.

What kind of heuristic design sufficiently performs this elimination? There are many but in this work we focus on rerandomization.


\subsection{The Rerandomization Design}\label{subsec:reandomization_design}

We denote the imbalance in $x$ as $B_x := \w^\top \x / (n / 2) = \xbar_T - \xbar_C$. A naive restricted design that eliminates poor $\w$'s goes as follows. (1) Realize a $\w$ from BCRD and measure $B^2_x$ and (2) if this $B^2_x$ is less than a predetermined threshold $a$, then retain $\w$ and run the experiment otherwise return to step (1).  Here, the variance-covariance matrix of the strategy $W$ will be dependent on both $\x$ and threshold $a$.

This \emph{rerandomization} heuristic dates to the inception of randomized experimentation. \citet[page 366]{Student1938} wrote that after an unlucky, highly imbalanced randomization, \qu{it would be pedantic to [run the experiment] with [an assignment] known beforehand to be likely to lead to a misleading conclusion}. His solution is for \qu{common sense [to prevail] and chance [be] invoked a second time}. Although this rerandomization design is classic, it has been rigorously investigated only recently \citep{Morgan2012}. However, selecting the optimal threshold wich we denote as $a_*$ \qu{remains an open problem} \cite[page 9162]{Li2018}. This threshold is a critical quantity because it controls the degree of randomness in the design. A miniscule $a$ will demand the optimal, deterministic design; a large $a$ would allow for complete randomization. Hence, solving this problem can bridge the intellectual gap between those that deterministically design experiments via optimization and those that design experiments via randomization.

\section{Methodology}\label{sec:methodology}

Herein, we provide procedures to locate this optimal threshold $a_*$ for the regression estimator under a realistic response model. For reasons that will become clear later, we leave an algorithm for the threshold in the context of the differences-in-means estimator for future work. Before we begin, we would like to reiterate that the rerandomization design features valid inference for any threshold (see Section~\ref{subsec:inference}). Thus, any specific $a_*$ that results from our algorithm will not affect the validity of the resulting inference.

We assume $p < n$ covariates measured for each subject and collect them into a row vector $\x_i := \bracks{x_{1,i}, \ldots, x_{p,i}}$. We denote $\X$ as the $n \times p$ matrix that stacks the $n$ vectors for each subject row-wise. Without loss of generality, we assume that each column in $\X$ is centered and scaled. We examine a more general response model (than we introduced in Equation~\ref{eq:simple_model}) linear in the observed covariates,

\bneqn\label{eq:realistic_response_model}
\y = \betaT\w + \X\bbeta + \z.
\eneqn

\noindent Here, $\bbeta$ is the vector of $p$ weights for each covariate as a column vector of length $p$. And $\z$ is the fixed contribution of unobserved measurements for the subjects as in the model of Equation~\ref{eq:simple_model} plus the misspecification errors of the true response function $f$ minus the linear component, i.e. $\bracks{f(\x_1) - \x_1 \bbeta \dots f(\x_n) - \x_n \bbeta}^\top$. Because each column in $\x$ is standardized, the linear component of this model does not require the additive intercept term.

In order to simplify our expressions, we note that the rerandomization design $W$ has the mirror property (Property~\ref{ass:mirror}) whereby the treatment group and control group can be swapped without changing the probability of the assignment.\footnote{Note that every design currently in wide use is mirror (e.g. BCRD, blocked designs, matched pair designs and of course, rerandomization, the design investigated herein). Further, a non-mirror design would be suspicious in the following way. Consider a partition of subjects into two subsets of treatments and controls. A non-mirror design would say you could not switch the treatments and controls subsets with equal probability.}\\

\begin{property}[Mirror Property]
\label{ass:mirror}
For all $\w \in \allocspace$, $\cprob{W = \w}{\X}$ = $\cprob{W = -\w}{\X}$. 
\end{property}

We now need to discuss how we gauge performance of our estimators (C4). We begin by employing the mean squared error, i.e. the weighted mean over the squared loss considering all allocations with $\X$ and $\z$ fixed. As an example, if $\prob{\w}$ is discrete uniform (such as in the case of BCRD and rerandomization), the expressions are computed via $|\allocspace|^{-1} \sum_{\w \in \allocspace} (\hat{\beta}(\w) - \betaT)^2$.

In \ref{app:mse_multivariate_simple} we show that in the randomization model, the MSE of $\betaThatDiffMeans$ is

\bneqn\label{eq:mse_simple_estimator}
\cmsesubnostr{\w}{\betaThatDiffMeans}{\z, \X; \bbeta} = \oneover{n^2} (\X\bbeta + \z)^\top \bSigmaw (\X\bbeta + \z)
\eneqn

\noindent and in \ref{app:mse_multivariate} we show that the MSE of $\betaThatLinRegr$ can be approximated to the third order as

\bneqn\label{eq:mse_linear_estimator}
\cmsesubnostr{\w}{\betaThatLinRegr}{\z, \X, \bbeta} \approx \oneover{n^2}  \z^\top \parens{\G + \frac{2}{n} \D} \z
\eneqn

\noindent where $\G :=(\I - \P) \bSigmaw (\I - \P)$, where $\P := \XXtXinvXt$ is the standard orthogonal projection matrix onto the column space of $\X$ and $\D := \expesub{\w}{\w \w^\top \P \w \w^\top}$, an expectation of a quartic form that cannot be simplified.\footnote{For intution about the behavior of Equation~\ref{eq:mse_linear_estimator}, see the Appendix's Table~\ref{tab:rand_model_estimators} that provides it for the special case of $p=1$.} The expression of Equation~\ref{eq:mse_linear_estimator} is independent of the unknown $\bbeta$ coefficients, a fact that will allow us to evaluate explicit designs in the following sections.

In theory, the optimal MSE rerandomization strategy is to locate $a_*$, the threshold corresponding to the minimum of Equation~\ref{eq:mse_simple_estimator} or \ref{eq:mse_linear_estimator} over $a$. As $a$ is varied, the determining matrix of the quadratic form varies as $\bSigmaw$ is a function of $\X$ and $a$. However, in practice this is impossible as $\z$ (and $\bbeta$ in the case of $\betaThatDiffMeans$) is unknown. We now discuss three criterions (C4) that remove this mathematical dependence on the fixed set of $\z$.

\subsection{Criterions of Design Optimality}

\subsubsection{The Minimax Design}\label{subsec:minimax_design}

If we assume nature to be adversarial, we can examine the MSE in the case of the supremum over the MSE, i.e. the worst possible finite vector of values $\z$. The optimal minimax strategy would then be the BCRD \citep[Theorem 2.1]{Kapelner2019} for $\betaThatDiffMeans$ and would asymptotically be BCRD for $\betaThatLinRegr$ (ibid, Appendix 6.11), corresponding to $a_* = \infty$. This result is not only trivial but would also correspond to the situation where $\z$ is aligned with one arbitrary direction in $\reals^n$, a punitively conservative and unrealistic situation.

\subsubsection{The Mean Unobserved Design}\label{subsec:mean_unobserved_design}

Can we consider the case of an \qu{average $\z$}? To do so, we must imagine many possible experimental datasets of size $n$ indexed by the elements of the support of $\Z$, assumed a continuous measure. Although this is incoherent in the randomization model (where $\z$ is fixed) and thus considering these many possible experimental datasets is tantamount to mixing the randomization model with the population model, we do so only in order to create a sensible criterion. The assumption of random $\Z$ is not used in any way during inference (Section~\ref{subsec:inference}). Since remind the reader that $\Z$ is exclusively a useful temporary mathematical device, we refer to it as the \qu{faux measure}. To obtain tractable results, we make the standard regression assumptions: we assume $\Z$ is mean centered (Assumption \ref{ass:random_z}) and the $n$ subjects' unobserved covariates are independent (Assumption \ref{ass:independence}). And, only to make our expressions simpler, we assume homoskedasticity in $\Z$ (Assumption \ref{ass:homo}). \\

\begin{assumption}[Faux Measure Mean Centeredness] \label{ass:random_z} $\cexpe{\Z}{\X, \bbeta, \w} = \zerovec_n$ i.e. linearity in the observed covariates.
\end{assumption}

\begin{assumption}[Faux Measure Independence] \label{ass:independence} $\cvar{\Z}{\X}$ is diagonal.
\end{assumption}

\begin{assumption}[Faux Measure Homoskedasticity] \label{ass:homo} $\cvar{Z_i}{\X} = \sigsq_z$ for all $i$.
\end{assumption}

\noindent In \ref{app:expe_mse_multivariate_simple}, we derive the mean criterion for $\betaThatDiffMeans$,

\bneqn\label{eq:expe_mse_diff_means_estimator}
\expesub{\z}{\cmsesubnostr{\w}{\betaThatDiffMeans}{\z, \X, \bbeta}} =  \frac{\sigsq_z}{n} + \oneover{n^2}\bbeta^\top \X^\top \bSigmaw \X \bbeta
\eneqn

\noindent and in \ref{app:mse_multivariate} we derive the mean criterion for $\betaThatLinRegr$ approximated to the third order as

\bneqn\label{eq:expe_mse_linear_estimator}
\expesub{\z}{\cmsesubnostr{\w}{\betaThatLinRegr}{\z, \X, \bbeta}} \approx  \frac{\sigsq_z}{n} + \frac{\sigsq_z}{n^2} \tr{\X_\perp^\top \bSigmaw \X_\perp}
\eneqn

\noindent where we denote $\X_\perp$ as the orthogonalized $\X$ matrix with columns $\x_{\perp_{\cdot 1}}, \ldots, \x_{\perp_{\cdot p}}$. 

The optimal design for $\betaThatDiffMeans$ would be the single vector $\w_*$ that minimizes $(\w^\top \X \bbeta)^2$. Note that this assumes knowledge of $\bbeta$ which is unknown in practice. Thus we leave this design problem for future work. The optimal design for $\betaThatLinRegr$ would be the single vector $\w_*$ that minimizes $\sum_{j=1}^p (\w^\top \x_{\perp_{\cdot j}})^2$ which does not depend upon $\bbeta$. This is similar to the optimal design under the population model setting as explained in \ref{app:population_model} which makes sense since we now have made the population model-like assumptions (\ref{ass:random_z}, \ref{ass:independence}, \ref{ass:homo}).

Since there are an exponential number of $\w$ vectors, the corresponding rerandomization threshold will be $a_* \approx 0$. Finding this vector is practically impossible as there is no known polynomial-time algorithm. Thus, this is not a practical way of selecting a rerandomization threshold. Even if it were, there would be inferential complications with such a deterministic design that are explained in Section~\ref{subsec:inference}.

\subsubsection{The Tail Unobserved Design}\label{subsec:tail_unobserved_design}

The supremum criterion is too conservative and the mean criterion does not incorporate the ruinous effect of vectors that can be near the supremum. Following \citet[Section 2.2.6]{Kapelner2019}, we consider a tail criterion for (C4) that we employ for the rest of this paper: the $q$th quantile of the MSE the estimator, i.e. $\text{Quantile}_z[\cmsesubnostr{\w}{\betaThat}{\z, \X, \bbeta}, ~q]$. This criterion gauges the average experimental error at the worst $1 - q$ percent of $\z$'s. For example, setting $q = 95\%$ would create a criterion that considers the \qu{5\% worst experiments}. 

As $q$ increases, the practitioner takes out insurance on unobserved covariates that are more and more improbable. As $q \rightarrow 100\%$, the only way to insure against every event is more randomness in $W$ (and more imbalance) as explained in Section~\ref{subsec:minimax_design}. As $q$ decreases towards the quantile corresponding to the mean, minimizing imbalance in $W$ becomes more important at the expense of less randomization as seen in Section~\ref{subsec:mean_unobserved_design}. Hence the tail provides a tradeoff between the two competing interests of optimality and randomization. We will understand the tradeoff explicitly in Section~\ref{subsec:est_Q_approx} but we first describe the algorithm.

\subsection{Algorithms to Optimize the Tail Criterion}\label{subsec:algorithms_opt_tail_criterion}


Our Algorithm~\ref{alg:master} (see \ref{app:algs}) is an exhaustive search to locate a minimal MSE design for $\betaThatLinRegr$. An analogous algorithm to locate the minimal MSE design for $\betaThatDiffMeans$ we leave to future work as it depends on $\bbeta$, an unknown quantity. We first outline two inputs to the algorithm besides the raw data $\X$ that are required. 

The algorithm takes as input a collection of assignments $\allocspacebase := \braces{\w_1, \ldots, \w_S}$ where $S$ is large, so that $\allocspacebase$ loosely approximates the full space $\allocspace$. Since \qu{rerandomization is simply a tool that allows us to draw from some predetermined set of acceptable randomizations} \citep[page 1267]{Morgan2012}, there is no theory we know of that specifies $\allocspacebase$. Thus, following \citet{Morgan2012} and \citet{Li2018}, we consider the initial set to be $S$ draws from BCRD. In practice, this decision can limit our algorithm's performance in certain situations that will become apparent in the simulation study in Section~\ref{sec:simulations} and strategies to mitigate these limitations are discussed in Section~\ref{sec:discussion}.

We then sort the assignments based on imbalance in $\X$. The imbalance calculation requires the second input: a metric of imbalance, of which our algorithm is agnostic. An imbalance metric $\imbal$ takes $\X$ and $\w$ as input and outputs a non-negative scalar. An output of zero indicates perfect balance between the treatment and control groups. One such metric is Mahalanobis distance, a collinearity adjusted squared sum of average differences in each dimension. This metric has nice properties; for instance it is an \qu{affinely invariant scalar measure} \citet[page 1271]{Morgan2012}. In our notation, it can be compactly expressed as  

\bneqn\label{eq:imbalance_compact}
\imbal(\X, \w) = \oneover{n}\w^\top \P \w.
\eneqn

\noindent We discuss other choices of metrics in Section~\ref{sec:discussion}. 

Let $\w_{(1)}$ be the vector with the smallest imbalance in the set and $\w_{(S)}$ be the largest, i.e. $\w_{(1)} := \argmax_{\w \in \allocspacebase} \braces{\imbal(\X, \w)}$. The closest vector to $\w_*$, the assignment that truly minimizes imbalance over the exponential number of theoretical vectors, in our small initial subset $\allocspacebase$ will be $\w_{(1)}$.

Beginning with $W \sim \{ \w_{(1)} $ w.p. 1 $\}$, we compute $Q$, the quantile tail criterion at $q$ for both the estimators. In order to calculate the quantile, we require the inverse CDF function in $\z$ for Equations~\ref{eq:mse_simple_estimator} and \ref{eq:mse_linear_estimator}. In general the closed form of the density functions are unknown asymptotically \citep[see e.g.][]{Gotze2002}. For both estimators, we present three procedures in the next three sections but first finish explaining the algorithm.

We then proceed to the strategy $W \sim \{ \w_{(1)}$ w.p. 1/2 and $\w_{(2)}$ w.p. 1/2$\}$ and recalculate $Q$ and then $W \sim \{ \w_{(1)}$ w.p. 1/3, $\w_{(2)}$ w.p. 1/3 and $\w_{(3)}$ w.p. 1/3$\}$ and recalculate $Q$. We repeat this procedure until the $S$th iteration where $W \sim U(\allocspacebase)$, i.e. the uniform discrete distribution which constitutes the finite approximation of BCRD. Over all iterations, there is a minimum quantile $Q_*$ corresponds to a an iteration number $s_* \leq S$, defining the approximate optimal strategy $\allocspace_* = \braces{\w_{(1)}, \dots, \w_{(s_*)}}$ and optimal threshold $a_*$ corresponding to the imbalance $\imbal(\X, \w_{(s_*)})$.

Our exhaustive search algorithm is computationally intensive but can be approximated by skip-stepping through the $S$ iterations at only a tiny loss of resolution. Also, from our empirical experience with hundreds of plot illustrations of $Q(a_*)$, we conjecture that the criterion is convex in $s$ but leave a proof for further work. We can provide a golden-section type line-search method which performs exponentially faster for those who wish to rely on this conjecture. 

We now discuss three strategies to compute $Q$ based on assumptions about the measure on $\Z$. Algorithm~\ref{alg:master} thus takes an argument that specifies one of three different tail computation functions found in Algorithm~\ref{alg:lr_tail_functions} for $\betaThatLinRegr$ (see \ref{app:algs}). These three we offer for convenience in order of the degree of practitioner knowledge. 

The first procedure (Section~\ref{subsec:est_Q_normality}) uses the usual OLS assumption of normality. The second procedure (Section~\ref{subsec:est_Q_approx}) allows the practitioner the flexibility to specify a departure from normality (summed up by a estimate of the excess kurtosis). This is useful if the practitioner happens to have prior knowledge about the nature of $\Z$. However, this use case would be rare but renders the software package more general and hence more powerful. The third procedure (Section~\ref{subsec:est_Q_provide}) requires full knowledge of the $\Z$ distribution. This full knowledge is completely unrealistic but may be useful in certain physical experimental settings and also is useful for comparison purposes during simulation to understand sensitivity. As we provide the practitioner with options, it is important to provide use-case recommendations and we do so in Section~\ref{subsec:recommendations}.

\subsubsection{Assume Normality and Use a CDF Approximation}\label{subsec:est_Q_normality}

In addition to assuming independence and homoskedasticity (Assumptions~\ref{ass:independence} and \ref{ass:homo}), we make the assumption that $\Z$ is Gaussian, then the distribution of the quadratic form of the MSE of $\betaThatLinRegr$ (Equation~\ref{eq:mse_linear_estimator}) is known explicitly. Since the optimal threshold $a_*$ will be invariant to shifts or scales of the MSE, we assume a standard Gaussian. Since the determining matrix is positive definite and symmetric, the distribution of this quadratic form is a standard result: a weighted sum of standard chi-squared distributions,

\bneqn\label{eq:weighted_sum_chi_squareds}
\cmsesubnostr{\w, \z}{\betaThatLinRegr}{\X, \bbeta} \approx  \Z^\top \parens{\G + \frac{2}{n} \D} \Z \sim \sum_{i=1}^n \lambda_i \chisq{1}
\eneqn

\noindent where the $\lambda_i$'s are the eigenvalues of  $\G + (2/n) \D$. Quantiles of this distribution are unknown in closed form but can be computed by numerical integration using the characteristic function. Fast approximations have been studied since the 1930's. \citet{Bodenham2016} compared many approximations on accuracy and computation time and recommend the Hall-Buckley-Eagleson method \citep{Buckley1988} which has error that is $O(n^{-1})$. This is the approximation used in our \texttt{R} package.


\subsubsection{Assume an Excess Kurtosis and Use a Double Approximation}\label{subsec:est_Q_approx}

The previous two strategies are weak in that they require an assumed distribution of $\Z$. However, any quantile can be expressed as

\bneqn\label{eq:hack_criterion}
Q := \expesub{\z}{\cmsesubnostr{\w}{\betaThat}{\z, \X, \bbeta}} + c \times \sesub{\z}{\cmsesubnostr{\w}{\betaThat}{\z, \X, \bbeta}}
\eneqn

\noindent where $q$ is one-to-one with $c$ given $\bSigmaw$ and $\se{\cdot}$ denotes the standard error of an expression with respect to the subscripted random variable. For example, if $q=95\%$ and the $\cmsesubnostr{\w}{\betaThat}{\z, \X, \bbeta}$ is normal, $c = 1.65$, the Gaussian quantile. The MSE for both estimators is conjectured to be asymptotically the form of Equation~\ref{eq:weighted_sum_chi_squareds} which is well-approximated by the normal distribution as long as (1) there are many non-zero eigenvalues relative to $n$ and (2) these eigenvalues are fairly uniformly distributed. These assumptions will be true as long as $W$ remains fairly random i.e. this approximation will work for rerandomization thresholds that are not too small and thus maintain randomness in $\allocspace$. The simulations of \citet[Section 3]{Kapelner2019} as well as much unshown simulation by its authors demonstrate that the Gaussian quantiles are approximately correct in many situations. This is the approximation employed in this section and in our \texttt{R} package.

In order to derive the standard error of the MSE's for both expressions, we need to assume that the $Z_i$'s have a finite fourth moment and that the third and fourth moments are independent of $\X$. \\

\begin{assumption}[Faux Measure has Finite Fourth Moment] \label{ass:finite_fourth} $\expe{Z_i^4} < \infty$ for all $i$.
\end{assumption}

\begin{assumption}[Faux Measure Independence of Higher Moments] \label{ass:high_moments_indep} $\expe{Z_i^3}$ and $\expe{Z_i^4}$ are independent of $\X$ for all $i$.
\end{assumption}

Although we do not have a procedure for finding the optimal threshold for $\betaThatDiffMeans$ due to its dependent on $\bbeta$, we derive the standard error of its MSE expression in order to gain intuition for the linear regression estimator we study later. In \ref{app:se_mse_multivariate_simple}, we derive the standard error for the MSE of the differences-in-means estimator,
 
\bneqn\label{eq:se_mse_diff_means_estimator}
&& \sesub{\z}{\cmsesubnostr{\w}{\betaThatDiffMeans}{\z, \X, \bbeta}}  = \nonumber\\
&&  \frac{\sigsq_z}{n^2} 
\tothepow{
	n \kappa_z + 2 \frobsq{\bSigmaw} + \frac{4}{\sigsq_z} \bbeta^\top \Xt \bSigmaw^2 \X\bbeta
}{\half}
\eneqn

\noindent where $\kappa_z$ is the excess kurtosis in $Z$ and $\frob{\cdot}$ denotes the Frobenius norm. Putting Equations~\ref{eq:expe_mse_diff_means_estimator} and \ref{eq:se_mse_diff_means_estimator} together and recalling that the optimal threshold $a_*$ will be invariant to shifts or scales of the MSE, we derive the proportional tail criterion objective function in Appendix~\ref{app:se_mse_multivariate_simple} and denote it $Q'$,

\bneqn\label{eq:tail_criterion_diff_means_estimator}
Q'_{\betaThatDiffMeans} =
\underbrace{\bbeta^\top \Xt \bSigmaw \X \bbeta}_{BAL_1}  +  \ingray{c\,\sigsq_z}
\Bigg(
	\ingray{n \kappa_z} + 
	\ingray{2} \underbrace{\frobsq{\bSigmaw}}_{RAND} + 
	\ingray{\frac{4}{\sigsq_z}} \underbrace{\bbeta^\top \Xt \bSigmaw^2 \X\bbeta}_{BAL_2}
\Bigg)^{\half}.
\eneqn

\noindent The quantities in grey are independent of the choice of design given the normal approximation. The $BAL$ terms measure observed imbalance. Over the range of possible $a_*$ values, the $BAL_1$ term ranges from $O(n)$ in the case of BCRD down to $O(n^3 2^{-n})$, i.e. effectively zero, in the case of the deterministic minimal imbalance \citep[Section 3.3]{Kallus2018}. 

For BCRD, $BAL_2 = n/(n-1) BAL_1 = O(n)$. We have the general upper bound $BAL_2 \leq n BAL_1$. The $RAND$ term measures design randomness and over the range of possible $a_*$ values, the $RAND$ term ranges from $\approx n$ in the case of BCRD all the way to $n^2$ in the case of the deterministic minimal imbalance. The tradeoff between imbalance and randomness is now clearly seen in tail criterion $Q'$.

In \ref{app:se_mse_multivariate}, we derive an approximation for the standard error for the MSE in the linear regression estimator,

\bneqn\label{eq:se_mse_linear_estimator}
&& \sesub{\z}{\cmsesubnostr{\w}{\betaThatLinRegr}{\z, \X, \bbeta}}  \approx   \nonumber\\
&& \frac{\sigsq_z}{n^2} \parens{2 \frobsq{(\I - \P) \bSigmaw} + \frac{8}{n}\tr{\G\D} + \frac{8}{n^2}\tr{\D^2} + \kappa_z SS}^{\half} 
\eneqn

\noindent where $SS := \sum_{i=1}^n \squared{g_{i,i} + 2 d_{i,i} / n}$. Putting Equations~\ref{eq:expe_mse_linear_estimator} and \ref{eq:se_mse_linear_estimator} together and dropping additive and multiplicative constants, we find an approximate proportional tail criterion objective function,

\bneqn\label{eq:tail_criterion_prime_linear_estimator}
Q'_{\betaThatLinRegr} &\approx& \underbrace{\tr{\X_\perp^\top \bSigmaw \X_\perp }}_{BAL} ~+~  \nonumber \\
&& \ingray{c} \,
\Bigg(
\ingray{2} \underbrace{\frobsq{(\I - \P) \bSigmaw}}_{RAND^2} + \ingray{8}\underbrace{\frac{\tr{\G\D}}{n}}_{\leq RAND \times BAL} + \ingray{8} \underbrace{\frac{\tr{\D^2}}{n^2}}_{\leq BAL^2} + \ingray{\kappa_z}
SS
\Bigg)^{\half}.
\eneqn

\noindent Once again, quantities in grey are independent of the choice of design and the tradeoff between imbalance and randomness\footnote{Note that the naive estimator for $\frobsq{(\I - \P) \bSigmaw}$, i.e. setting $\bSigmaw$ equal to the estimate $\bSigmawhat = \oneover{S} \sum_{s=1}^S \w_s \w_s^\top$ and then adding all the squared entries of $(\I - \P)\bSigmawhat$, is biased upwards by Jensen's inequality. If this naive estimate were to be used in our computations, it would erroneously place more weight on the RAND term, pushing the optimal design more towards BCRD and thus increasing $a_*$. To mitigate this, we derive a modified estimator computed by de-biasing the naive estimator (see \ref{app:unbiased_frobsq_const}). This de-biased estimator we found to be unstable at small $S$. Future work will investigate this phenomenon further. Since the bias in our current expression is only $O(S^{-1})$, our designs will be conservative only for small $S$ which is irrelevant as we recommend $S$ to be in the tens of thousands when using our procedures.} is clearly observed as in the case for $\betaThatDiffMeans$. The inequalities underbraced above are proven in \ref{app:se_mse_multivariate}.

\subsubsection{Provide a Distribution Explicitly}\label{subsec:est_Q_provide}

If the distribution of $\Z$ is provided explicitly, then in each iteration of Algorithm~\ref{alg:master}, the empirical quantile $q$ of the MSE of $\betaThatLinRegr$ (Equation~\ref{eq:mse_linear_estimator}) can be approximated as the mean over many draws of $\z$. This will not only be computationally slow, but the average MSE at each $a$ will be noisy, and the criterion will have to be smoothed e.g. via cubic smoothing splines \citep[Chapter 2.2]{Hastie1990}. The optimal threshold $a_*$ will be invariant to shifts or scales of the MSE and thus the $\Z$ distribution can be standardized without changing the result. Note that assuming independence and homoskedasticity (Assumptions~\ref{ass:independence} and \ref{ass:homo}) are unnecessary in this section. 

\subsection{Frequentist Inference in Our Designs}\label{subsec:inference}

To run the experiment, we first draw one $\w_{exp}$ from our optimal threshold rerandomization strategy $W$ and these treatments are administered to the subjects. Then, responses for the treatment group $Y_{T,1}, \ldots, Y_{T,n/2}$ and for the control group $Y_{C,1}, \ldots, Y_{C,n/2}$ are collected. Using these responses, we wish to test theories about $\beta_T$ and produce a confidence interval for $\beta_T$.

Under the randomization model, the only randomness is in $\w$ and thus a randomization test is most appropriate; this was Fisher's \qu{reasoned basis} for inference. The randomization test \qu{can incorporate whatever rerandomization procedure was used, will preserve the significance level of the test and works for any estimator} \citep[p. 1268]{Morgan2012}. This test requires a sharp null hypothesis whereby all subjects' response will be exactly equal under both treatment and control conditions.

First, using the experimental $\w_{exp}$ we compute $\hat{\beta}_{exp}$ via linear regression. A null distribution can then created by using every other $\w \in \allocspace_*$, the curated set of assignments thresholded by the optimal $a_*$, and computing the same estimate (since these estimates were computed by labeling responses to treatment and control in a way unrelated to the design used in the actual experiment). \qu{One should analyze as one designs} \citep[p. 105]{Rosenberger2016}, so that replicates can only be drawn from the same restricted allocation set where $\w_{exp}$ originated. 

If $\allocspace_*$ is large, the null distribution can be approximated with $R < \abss{\allocspace_*}$ draws from $\allocspace_*$ where the value of $R$ controls the precision of the $p$-value. Each of the $R$ replicates involves drawing one $\w \in \allocspace$ and recomputing the estimates. 

A properly-sized but $\alpha$-level two-sided test can be run by assessing if the experimental $\hat{\beta}_{exp}$ falls in the retainment region bounded by the $\alpha/2$ and the $1-\alpha/2$ quantiles of the replicate set $\hat{\beta}_1, \ldots, \hat{\beta}_R$. By the duality of confidence intervals and hypothesis tests, one can alter the sharp null by adding (or subtracting) $\delta$ and rerun the test. The set of values of $\delta$ where the test rejects constitutes an approximate $1-\alpha$ confidence interval for the average treatment effect. This is computationally expensive, but efficient algorithms exist \citep[e.g.][]{Garthwaite1996}.

If the optimal threshold $a_*$ is small, our design \qu{must leave enough acceptable randomizations to perform a randomization test} \citep[page 1268]{Morgan2012} and there may not be enough assignments out of the total $S$ that satisfy this threshold.  The smallest imbalance values are the left tail of the distribution $\braces{\imbal(\X, \w)\,:\,\w \in \allocspacebase}$ which are $O_p(p S^{-2/p})$ in the case of the $\X$ being sampled from an iid Gaussian generating process and $\imbal$ is the Mahalanobis distance. \citet[Section 3.3]{Kallus2018} proved that the imbalance for the optimal assignment in this case is $\imbal(\X, \w_*) = O(n 2^{-2n})$. Thus, BCRD will not be able to locate vectors far into this left tail in a reasonable amount of time.

Additionally, estimating the terms in our criterion $Q'$ of Equation~\ref{eq:tail_criterion_prime_linear_estimator} may suffer from smaller and smaller sets of assignments. The RAND term for instance relies on an accurate estimate of a function of the $n \times n$ variance-covariance matrix $\bSigmaw$ and the $n \times n$ matrix $\D$. These matrices have an $O(n^2)$ number of parameters and require many samples for high accuracy and a bias correction.


\section{Simulations for the Linear Regression Estimator}\label{sec:simulations}

\subsection{The Three Strategies in Different Settings}\label{subsec:three_strategies}

\begin{figure}[htp]
\centering
\includegraphics[width=5in]{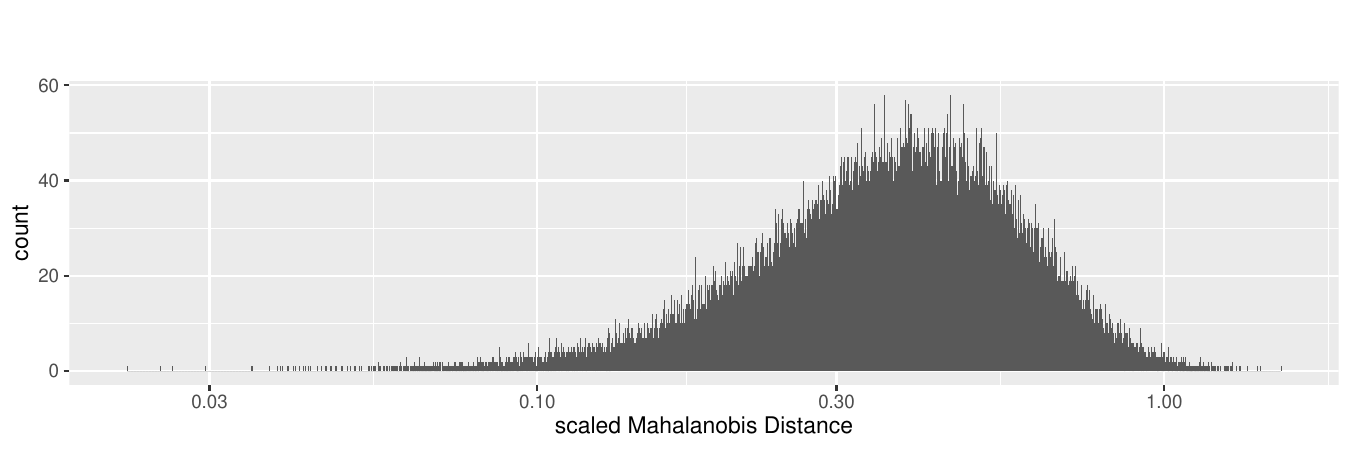}
\caption{Imbalances in $\allocspacebase$ as measured by the Mahalanobis distance save constants.}
\label{fig:baseline_imbalances}
\end{figure}

We first simulate a case where $n=100$ and $p=10$ where the entries of $\X$ are standard normal realizations. We begin with $\allocspacebase$ of size $S=25,000$ sampled from BCRD. The baseline imbalances are given in Figure~\ref{fig:baseline_imbalances}. It is the job of our algorithms to threshold this distribution at $a_*$ to optimize the tail of estimator MSE.

\begin{figure}[t]
\centering
\includegraphics[width=5.3in]{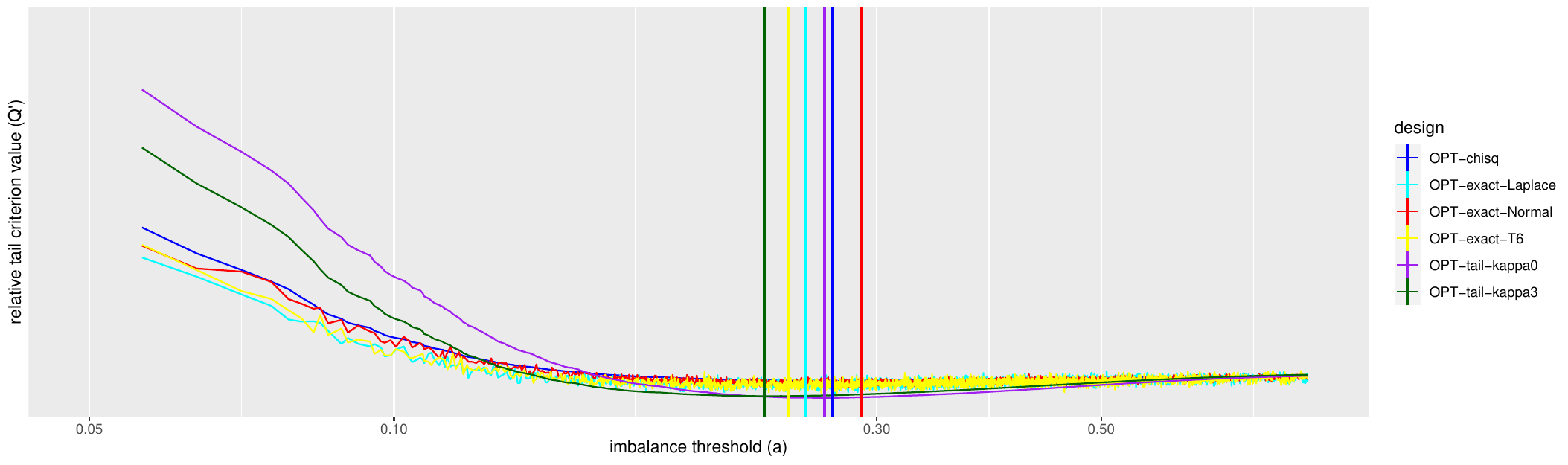}
\caption{Relative tail criterion ($Q$ save additive and multiplicative constants) for $q=95\%$ for many settings and strategies plotted versus $a$ on a log scale. The vertical line indicates the optimal threshold $a_*$ for each strategy. For the purpose of illustration, all tail criterions were set to 1 at their 99\%ile values so that the six curves coincide in order to clearly visualize differences.}
\label{fig:all_methods_norm_laplace_t2}
\end{figure}

For the linear estimator, we run Algorithm~\ref{alg:master} to find the optimal threshold for many settings and plot the 95\% tail criterion over $a$ in Figure~\ref{fig:all_methods_norm_laplace_t2}. We simulate three settings where the faux measure is normal: OPT-chisq denotes the normal approximation of Section~\ref{subsec:est_Q_normality}, OPT-tail-kappa0 denotes the tail approximation of Section~\ref{subsec:est_Q_approx} with $\kappa_z = 0$ and OPT-exact-Normal denotes the exact strategy of Section~\ref{subsec:est_Q_provide} simulated with 1,000 standard normal realizations for each subset $\allocspace_0$ and smoothed. We also simulated three settings where the faux measure is leptokurtic: OPT-exact-Laplace denotes the exact strategy for $\Z$ being the standard Laplace distribution simulated with 1,000 realizations for each subset $\allocspace_0$ and smoothed, OPT-exact-T6 denotes the exact strategy for $\Z$ being the Students' $t$ distribution with six degrees of freedom with 1,000 realizations for each subset $\allocspace_0$ and smoothed and OPT-tail-kappa3 denotes the tail approximation with $\kappa_z = 3$. Both the Laplace distribution and Students' $t$ distribution with six degrees of freedom have excess kurtosis of 3. Thus, the tail approximation in OPT-tail-kappa3 sets $\kappa_z = 3$ to test our tail approximation procedure's ability to approximate this degree of leptokurtosis.

There are many observations made from this simulation. First note that for all plots of exact distributions the criterion is essentially flat until $a < 0.20$. At these thresholds, there is too much restriction which will be detrimental. Any $a_*$ design where $a > 0.20$ (a large range of potential thresholds) will be about equal and will not effect tail MSE performance as will be shown in Section~\ref{subsec:visualizing_tail}.

Second, all $a_*$ values returned by any of the six procedure-design settings is about the same and hence the number of vectors $\abss{\allocspace_*}$ will also be about the same. Their values can be read from the vertical lines in Figure~\ref{fig:all_methods_norm_laplace_t2} and are as follows: OPT-chisq has $a_* = 0.271$, OPT-tail-kappa0 has $a_* = 0.266$, OPT-exact-Normal has $a_* = 0.289$, OPT-exact-Laplace has $a_* = 0.255$, OPT-exact-T6 has $a_* = 0.245$ and OPT-tail-kappa3 has $a_* = 0.232$. This indicates a great degree of robustness to the assumptions on $\Z$ and choice of procedures implying unambiguous advice to a practitioner (see Section~\ref{subsec:recommendations}).

%

\subsection{Visualizing the Optimal MSE Tail}\label{subsec:visualizing_tail}

We now simulate response models of Equation~\ref{eq:realistic_response_model} under different scenarios to demonstrate that our algorithm finds the optimal threshold for a tail quantile. We retain the same $\X$ from before with $n=100$ and $p=10$. To draw responses, we set $\beta_T = 1$ and draw $\bbeta$ from $\multnormnot{p}{\zerovec_p}{\I_p}$. The entire simulation is fixed upon these values. We then draw 5,000 different $\z$'s from $\multnormnot{n}{\zerovec_n}{\sigsq_z \I_n}$ where $\sigsq_z = 3$ so that to $R^2_{\X} \approx 26\%$, emulating a typical clinical case where the covariates are only modestly explanatory of the outcome measure. 

We employ seven design strategies: BCRD, balanced complete randomization with $S=25,000$ vectors; DET, the deterministic design of the best vector out of the $S$ vectors; GOOD, a naive selection of the 0.5\% least imbalanced vectors of all $S$;  BAD, a naive selection of the 0.5\% most imbalanced vectors of all $S$ and our three proposed procedures: OPT-chisq, the optimal design according to the weighted chi-squared approximation algorithm of Section~\ref{subsec:est_Q_normality}; OPT-tail, the optimal design according to the tail criterion approximation algorithm of Section~\ref{subsec:est_Q_approx} and OPT-exact, the optimal design when simulating $\Z$'s exact distribution algorithm of Section~\ref{subsec:est_Q_provide}.

\begin{figure}[h]
\centering
\begin{subfigure}[b]{\textwidth}
\includegraphics[width=6in]{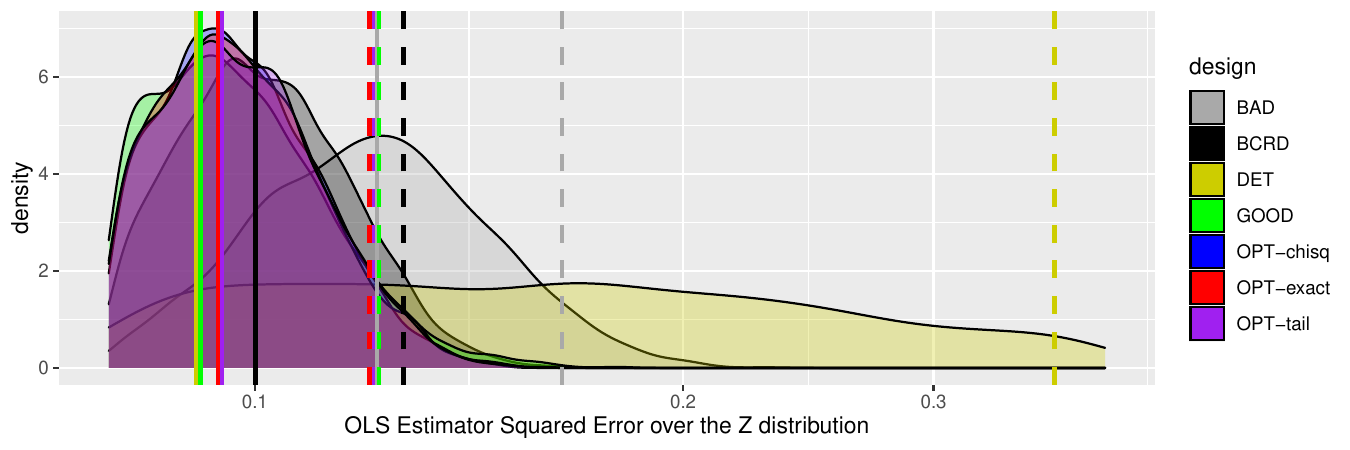}
\caption{All seven designs}
\label{fig:mse_different_strategies_all}
\end{subfigure}

\begin{subfigure}[b]{\textwidth}
\includegraphics[width=6in]{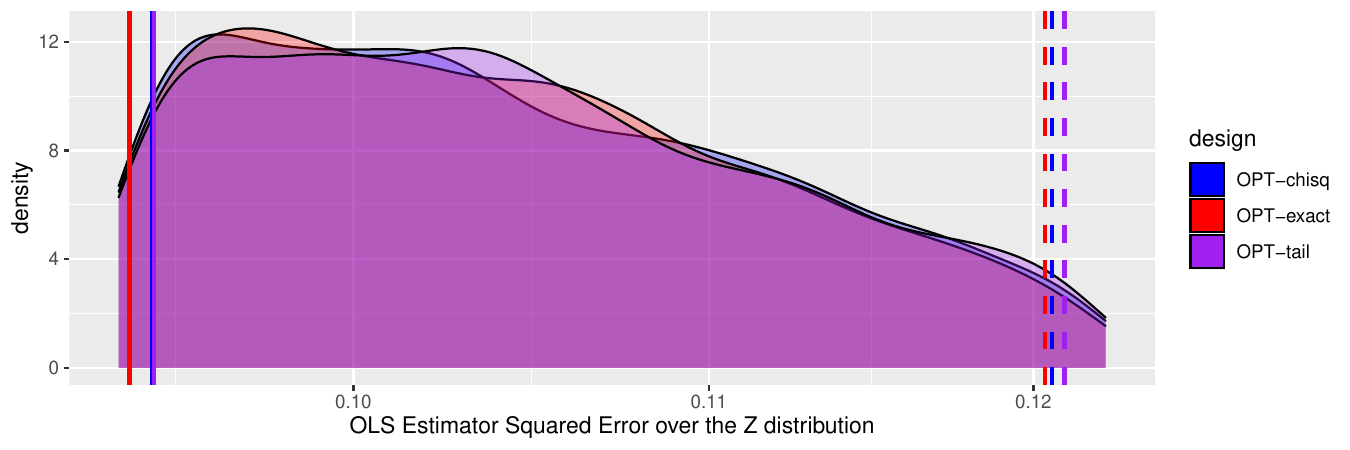}
\caption{Closeup of the three proposed rerandomization designs in this paper}
\label{fig:mse_different_strategies_three}
\end{subfigure}

\caption{Empirical MSE densities for the different design procedures. The solid vertical lines are the average MSE and the dotted vertical lines are the empirical 95\% quantiles.}
\label{fig:mse_different_strategies}
\end{figure}

For each design and for each of the 5,000 draws of $\z$, we draw 1,000 different $\w$ vectors from the BCRD, OPT-chisq, OPT-tail and OPT-exact designs, all 0.5\%$S$ vectors in the GOOD and BAD designs and the one optimal vector from the DET design, $\w_{(1)}$. We then compute $\y$ via Equation~\ref{eq:realistic_response_model} and $\betaThatLinRegr$ from $\X$, $\w$ and $\y$ and then we record the squared error $(\betaThatLinRegr - \betaT)^2$. The average over all $\w$'s is recorded as an $\cmsesubnostr{\w}{\betaThatLinRegr}{\z, \X, \bbeta}$. Over the 5,000 draws of $\z$, the MSE density is estimated. Figure~\ref{fig:mse_different_strategies_all} shows the MSE density estimates for all design strategies. Then, the expectation of the MSE over $\Z$ is estimated (the vertical solid lines) as well as the 95\% empirical quantile estimates the tail criterion (the vertical dashed lines).

There are many observations from this figure. First, as Equation~\ref{eq:expe_mse_linear_estimator} predicts, DET has the lowest expected MSE as shown as the solid yellow line. However, note the shape of the MSE density for DET with its long right tail. The tail shape can be understood by examining the first term in the MSE expression (Equation~\ref{eq:mse_linear_estimator}) that can be written as $\oneover{n^2} \z_\perp^\top \bSigmaw \z_\perp$ where $\z_\perp$ denotes the component of $\z$ orthogonal to the column space of $\X$. In the case of DET, the quadratic form becomes $(\z_\perp^\top \w_{(1)})^2$. This implies that for $\z$'s that interact poorly with the single, lowest imbalance allocation $\w_{(1)}$, there is a lot of error. The dotted yellow line is its 95\%ile which is more than twice as large as the maximum of the visible tails of the other designs' MSE densities (save BAD).  

Second, the OPT-chisq, OPT-tail and OPT-exact designs (the best 5,881, 5,601 and 6,901 vectors out of the original $S=25,000$ respectively) provide the best insurance against the 95\% worst MSE (the blue, red and purple dotted lines respectively) and are the lowest 95\%ile among all the seven strategies. They also perform the best for the mean MSE and only nominally larger than the mean MSE for DET (which again features a long tail of disastrous realizations). 

Third, the OPT-chisq, OPT-tail and OPT-exact designs perform similarly (Figure~\ref{fig:mse_different_strategies_three}). More important is that OPT-chisq and OPT-tail produce similar answers as these are the two choices a practitioner has when implementing our proposed designs (we elaborate on recommendations in Section~\ref{subsec:recommendations}). One reason why the OPT-tail may feature a slight performance penalty in error when compared to the OPT-chisq design is that it is designed to be robust to other distributions besides the Gaussian (meaning it will be closer to BCRD, have a larger rerandomization threshold and more vectors in $\allocspace_*$). Additionally, this procedure also has more variability due to the double approximation explained in Section~\ref{subsec:est_Q_approx}.

Fourth, there is a lot of leeway in finding the optimal design. The GOOD strategy of the top 0.5\% assignments performs about the same as all of our OPT designs (but our OPT designs do indeed provide a lower quantile). This is due to the flatness of the tail criterion among designs that are neither too random or too deterministic (see discussion below Figure~\ref{fig:all_methods_norm_laplace_t2} about the range where $a > 0.20$).

Fifth, BCRD performs somewhat worse, but not terrible when compared to OPT. This is a testament to the advantage of the OLS estimator --- the contribution of a priori imbalance is attenuated by a factor of $1/n^2$, a rapid rate of vanishing (Equation~\ref{eq:expe_mse_linear_estimator}). We will see in the next section that OLS cannot rest on its laurels in the case of larger $p$.

Sixth, the BAD strategy (the worst 0.5\% assignments) performs much worse in both the expectation and the tail criterions. This illustrates a case of covariate imbalance that is so lopsided that even OLS cannot fix it.

\subsection{Optimal Rerandomization Threshold Dependence on $p$}\label{subsec:p_dependence}

We now investigate how the optimal threshold $a_*$ depends on the number of observed covariates $p$. We fix $n=200$, $S=25,000$ and iteratively add more columns to $\X$ so that $p \in \braces{1, 2,  \ldots, 199}$ where new column entries are standard normal realizations. For each $p$, we compute the optimal threshold via the weighted chi-squared approximation algorithm  of Section~\ref{subsec:est_Q_normality} and record $\abss{\allocspace_*(p)} / S$, i.e. the fraction of vectors in the original set of $S$ that meet this threshold.

\begin{figure}[t]
    \centering
    	\includegraphics[width=5in]{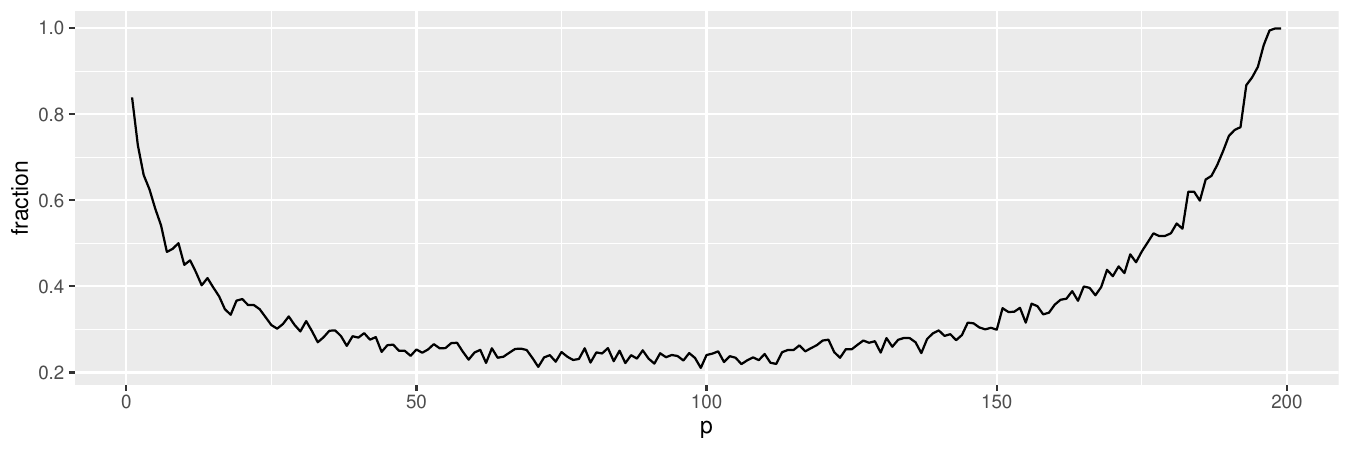}
    \caption{An investigation into the optimal threshold's dependence on the number of covariates: the fraction of base vectors less than $a_*$ versus the number of covariates $p$.}
	\label{fig:a_star_by_p_a_and_b}
\end{figure}

Figure~\ref{fig:a_star_by_p_a_and_b} shows that the fraction of acceptable vectors seems to be converging to zero when $p < \half n$. This portion of the illustration is consistent with our intuition: as there are more observed covariates, a priori restrictions on the randomization to decrease imbalance become more important to the resultant MSE tail (even when employing the linear adjustment using OLS). A decrease in imbalance means a decrease in the level of randomness and this seems to be well-justified. 

The second portion of the illustration (i.e. when $p > \half n$) displays the fraction of acceptable vectors increasing and thus $Q$, the tail of the MSE, must be increasing. As $p \rightarrow n$, then $\P \rightarrow \I_n$ implying that $\imbal(\X, \w) \rightarrow 1$ (Equation~\ref{eq:imbalance_compact}) meaning there is a large imbalance between the treatment groups with very little variability meaning that all risk is in the unmeasured confounders. To illustrate this point more clearly, $\P \rightarrow \I_n$ implies that $\G \rightarrow \bv{0}_n$ and $\D \rightarrow n\bSigmaw$ and thus $\cmsesubnostr{\w}{\betaThatLinRegr}{\z, \X, \bbeta} \rightarrow \frac{2}{n^2}\z^\top\bSigmaw\z$ (Equation~\ref{eq:mse_linear_estimator}). Because there is no longer much variability in $\imbal(\X, \w)$ the expectation of the MSE over $\Z$ (Equation~\ref{eq:expe_mse_linear_estimator}) will be relatively unaffected by the choice of $a_*$. Thus, the quantile is almost completely controlled by the variability in MSE over $\Z$. This standard deviation expression (Equation~\ref{eq:se_mse_linear_estimator}) converges to a constant times $(\text{tr}[\bSigmaw^2])^{1/2}$ which is minimized by employing BCRD i.e. $a_* \rightarrow \infty$ (with only a nominal effect of $\kappa_z$). Luckily, most randomized experiments and clinical trials do not feature very many informative control variables relative to number of observations. 

One final point is that the relationship does not appear to be smooth. Theoretically it should be smooth and convex and the observed jaggedness is likely due to the $O(n^{-1})$ numerical error in the Hall-Buckley-Eagleson method.

\subsection{Designing an Educational Intervention Experiment}\label{sec:clinical_trial}

The Tennessee Student Teacher Achievement Ratio (STAR) randomized experiment measured the effect of class size on student standardized test scores \citep{Word1990}. Similar to \citet[Section 5]{Kallus2018b}, we used this data set to validate our methodology using simulation on subsamples.

We focused on two of the experimental conditions: small classes versus regular-sized classes. The outcome metric we used was the average of the reading, math, language and listening standardized aptitude tests. Covariates we included and their data types were student gender (binary and 1 for male), race (binary and 1 for African American), class size (integer), whether they received free lunch (binary), birthdate (continuous), rural home address (binary), suburban home address (binary) and inner city home address (binary) for a total of $p=8$ control covariates. The experiment monitored students in kindergarten through third grade. We only used the outcomes in the final year. After dropping students with missing data, we were left with a total of $n = 3,685$ observations.

We then paralleled the MSE measurement simulation of Section~\ref{subsec:visualizing_tail} for BCRD, DET, OPT-chisq and OPT-tail (with $\kappa_z = 0$). We did not run OPT-exact as this procedure is not realistic as the practitioner would not have knowledge of the exact distribution of $\Z$ in a social science experiment setting. 

As in the previous simulation, in order to measure MSE by averaging squared errors over $\w$ as a function of $\z$, we need a way to vary both $\w$ and $\z$. Since the experimental data has one fixed known $\w$ and one fixed unknown $\z$, we employ a parametric bootstrap. We model the underlying data generating process assuming a response model additive in the conditional expectation, PATE and residuals:

\bneqn\label{eq:pate_model}
\y = f(\x_{\cdot 1}, \ldots, \x_{\cdot 8}) + \betaT \w + \z.
\eneqn

\noindent We estimated $\betaT = 3.056$ on the whole dataset and this was employed as baseline truth. We then subtracted $\betaT$ from all observations' responses which were assigned the treatment and added $\betaT$ to all observations' responses which were assigned the control. We then used a Random Forest model \citep{Breiman2001} to flexibly fit non-linearities and interactions within $f(\x_{\cdot 1}, \ldots, \x_{\cdot 8})$, the conditional expectation function, by fitting the $\betaT$-adjusted responses with the covariates. We recorded the predictions of $f(\x)$ for all $\x$ denoted ${\bv{\hat{ f}}}$ which were computed out-of-bag (in order to not use overfit predictions). We then computed the residuals as $\z := \y - \bv{\hat{ f}}$. 

The parametric bootstrap then works as follows: we sample $n$ rows from $\X$ with replacement  along with their corresponding conditional expectation function estimates. We then draw the $n$ residuals independently of the $n$ rows, using a sample with replacement $\z_{rep}$ from the set of all $\z$. This residual sampling procedure assumes homoskedasticity in the faux measure $\Z$ and the experimental subjects are assumed independent.

For this fixed $\z_{rep}$ we can compare the experimental designs by drawing $\w$ from the design, computing the response value via Equation~\ref{eq:pate_model}, computing the estimate $\betaThatLinRegr$, computing the squared error $(\betaThatLinRegr - \betaT)^2$ and averaging these squared errors over the $\w$'s to estimate $\cmsesubnostr{\w}{\betaThatLinRegr}{\z, \X, \bbeta}$. The variability in this MSE is then assessed by doing many iterations of these steps for different $\z_{rep}$'s.

\begin{figure}[h]
\centering
\includegraphics[width=6in]{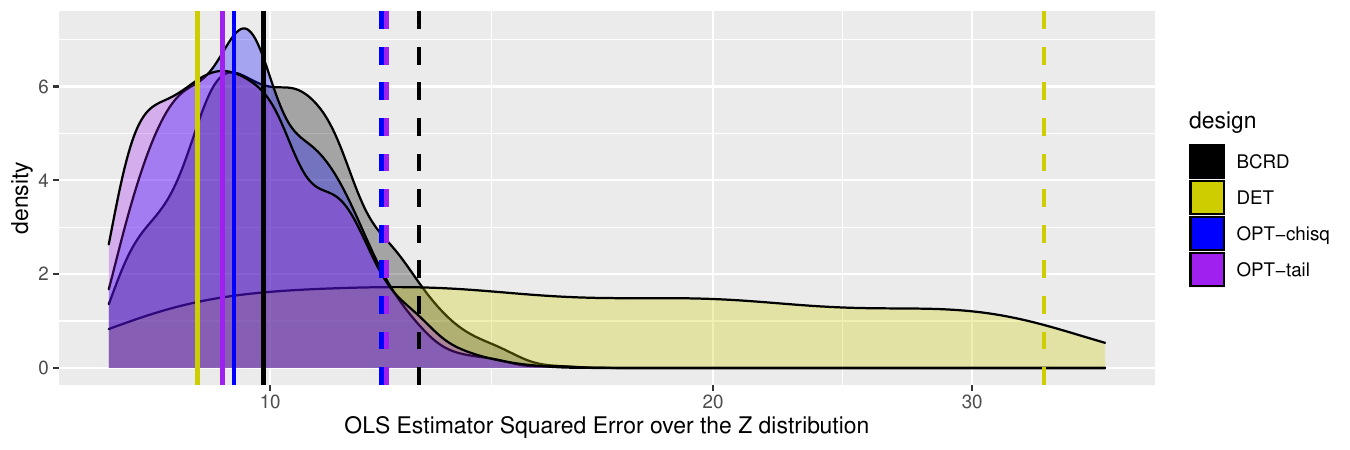}
\caption{Empirical MSE densities for the different design procedures. The solid vertical lines are the average MSE and the dotted vertical lines are the empirical 95\% quantiles.}
\label{fig:star_experiment}
\end{figure}

In our simulation, we used $n=100$ students and $S = 25,000$. Figure~\ref{fig:star_experiment} parallels Figure~\ref{fig:mse_different_strategies} by illustrating the density of these conditional MSE's over all 1,000 $\z_{rep}$'s.

Our results indicate that the OPT-chisq and OPT-tail rerandomization designs outperform BCRD in the mean MSE by 4.6\% and 6.2\% respectively and in the 95\% tail of the MSE by 5.7\% and 4.9\%. The tail is the target of our optimization procedure and thus a performance boost is expected but the performance gain in the mean MSE comes for free because we employed restricted randomization providing a boost even with an OLS estimator. Once again DET slightly beats the OPT designs in the mean but to realize that tiny benefit, one pays a big price: the 95\% tail MSE event is almost 300\% larger. Note that other samples of students resulted in illustrations that were not appreciably different.

This simulation also further vindicates the idea that the OPT-chisq and OPT-tail designs perform similarly (the third observation on Figure~\ref{fig:mse_different_strategies_three}). This is important because the OPT-chisq and OPT-tail are the two choices a practitioner has at the design stage.

These results are further expected as sample size, number of covariates, as well as the $R^2_{\X}$ (which was found to be $22\%$ when averaged over 1,000 samples of size $n$) parallel the conditions simulated before in Section~\ref{subsec:visualizing_tail}. As mentioned before, this gain will be attenuated as sample size increases due to the OLS estimator's ability to adjust covariate imbalance a posteriori obviating the need to restrict the randomizations to improve imbalance a priori.

\section{Discussion}\label{sec:discussion}

We have introduced an algorithm that can provide an optimal rerandomization threshold for the randomization model when inferring the average treatment effect using the covariate-adjusted linear regression estimator when insurance on high error is sought. The expressions we derived and our simulations confirm that relatively little rerandomization restriction is needed in the prevalent practical setting where $p$ is small relative to a large $n$. This underscores the robustness of complete randomization coupled with linear adjustment as a design-estimation approach in experimentation. However, experimental settings with small sample size are likely to benefit from our approach.

One can argue that we have replaced explicitly picking a value of $a_*$ with instead picking a value of $q$ in $\text{Quantile}_z[\cmsesubnostr{\w}{\betaThat}{\z, \X, \bbeta}, ~q]$, the quantile in our tail criterion. We would argue that one does not understand the scale of $a_*$ but the $q$ is much more interpretable: it is taking out insurance on bad experiments. Further, the range of reasonable $q$ choices is small, likely at most between 80-99\%.

Our work leaves many unanswered questions. First, there are other imbalance metrics and the choice of Mahalanobis distance seems arbitrary or one of convenience especially when real-world response models are non-linear. Recently, \citet[Section 2.3]{Kallus2018} proved that the optimal imbalance metric is dependent on the choice of norm for the response model. For instance, if the norm is Lipschitz, then the optimal $\imbal$ would capture the deviation from a pair-matching allocation, such as pairwise Mahalanobis distance (ibid, Theorem 4). \citet{Krieger2020} propose a design that uses both rerandomization and matching together. If the norm is the sup norm, then the optimal $\imbal$ is one that captures deviation from an incomplete blocking design (ibid, Theorem 3). If the squared norm is $n^{-1} \bbeta^\top \XtXinv \bbeta$, i.e. a measure of the signal in the linear model, then the optimal $\imbal$ is the Mahalanobis distance (ibid, Section 2.3.3). If the response function is represented by a reproducing Kernel Hilbert Space, then the optimal $\imbal$ would be $ \w^\top \K \w$ where $\K$ is the $n \times n$ \emph{gram matrix}, a matrix whose $i,j$ entries tally the kernel distance from $\x_i$ to $\x_j$ (ibid, Equation 4.2). These kernel spaces can be infinite in dimension and thereby can model non-parametric response functions. The exponential kernel with kernel distance function $\expnostr{\x_i^\top\x_j}$ can be seen as modeling all polynomials (with infinite degree) and the Gaussian kernel with kernel distance function $\expnostr{-\normsq{\x_i - \x_j}}$ is especially flexible (the two examples in ibid, Section 2.4).

Since the true model is unknown, one would want to employ an $\imbal$ sufficiently flexible to not suffer poor performance if the model deviates from their arbitrary choice. Thus, the Gaussian kernel is recommended in practice (ibid, Section 6). Our software contains options for a wide variety of choices for $\imbal$.

%

Additionally, as mentioned in Section~\ref{subsec:algorithms_opt_tail_criterion}, there is no theory we know of that specifies how $\allocspacebase$ should be sampled from the full space $\allocspace$. We have employed BCRD here but there may be other choices for instance those that provide more vectors with exponentially smaller imbalance. For example, \citet[Algorithm 1]{Krieger2019} is a greedy heuristic that begins with BCRD and in each iteration finds the switch of treatment-control subject pair that is optimal until no more minimization is possible. This algorithm finds assignments that are $O_p(p n^{-1 - 2/p})$. A nice property of this greedy heuristic is that it converges quickly requiring very few switches and thus their assignments are nearly as random as BCRD. Thus, including assignments from their algorithm are unlikely to affect the $RAND$ terms but greatly decrease the $BAL$ terms in our criterion $Q'$. Since the imbalance of vectors of BCRD is known, the greedy pair-switching algorithm can be used for stratified sampling for imbalance distributions upper bounded at small $a_*$. Our algorithm would proceed identically, but early simulation suggests the optimal designs can change drastically. This we believe is the most pressing open question for future research.

\subsection{Recommendations for the Practitioner}\label{subsec:recommendations}

We provided two procedures to locate $a_*$ based on the assumptions of the $\Z$ distribution: the first is to assume normality and the second is to specify a departure from normality. (Our third procedure requires the unrealistic full specification of the $\Z$ distribution and thus we omit it in this discussion). Overall, our procedures seem to have a high degree of robustness: differing assumptions about $\Z$ will only result in minor effects on the estimation error. Thus, we believe that using the first procedure (the normality assumption) should be used in practice unless the practitioner feels strongly enough that the errors (plus misspecification) have significantly thinner or fatter tails than a Gaussian in which case the second procedure could be used. 

Using the second procedure in place of the first when the normality assumption is justified only incurs a slight cost (as seen when comparing the red and purple vertical lines of Figures~\ref{fig:all_methods_norm_laplace_t2}, \ref{fig:mse_different_strategies_three} and \ref{fig:star_experiment}). The difference between both procedures will likely only be large under high $R^2$ settings which are not the norm at least in clinical trials. 

Regardless of the procedure employed or the value of $q$ chosen, we reiterate that our rerandomization design features valid inference (an unbiased estimator and properly sized hypothesis tests) for any value of $a$ (see Section~\ref{subsec:inference}). 

\subsection*{Replication}

The figures and values given in Section~\ref{sec:simulations} can be duplicated by running the code found in \url{https://github.com/kapelner/OptimalRerandExpDesigns/blob/master/JSPI_paper_figures_and_data}. 

\section*{Acknowledgements}

We thank Nathan Kallus and Uri Shalit for helpful discussions. This research was supported by Grant No. 2018112 from the United States-Israel Binational Science Foundation (BSF).

\bibliographystyle{model2-names}
\bibliography{refs}


\pagebreak
\appendix
\section{Appendix}

\subsection{The Population Model Perspective}\label{app:population_model}

In the population model \citep[Chapter 6.2]{Rosenberger2016}, subjects in the treatment group are sampled at random from an infinitely-sized superpopulation and thus the responses in the treatment group are considered independent and identically distributed with density $f_{Y}(y~|~\theta = \betaT)$. Subjects in the control group are sampled analogously from a different superpopulation density $f_{Y}(y~|~\theta = - \betaT)$ where $\theta$ denotes one of many possible parameters. 

In the simple model of Equation~\ref{eq:simple_model}, the population model can be created via the assumption that $\z$ is a random draw from $\Z$ such that $\z := \y - \cexpenostr{\Y}{\x, \w}$ and its entries are independent. The law of iterated expectation yields that the distribution of the $Z's$ are mean-centered. Each experiment gets one draw from $\z$ which is usually referred to as \qu{noise} or \qu{measurement error}. To make our expressions simpler, we also assume homoskedasticity so that each of the $Z_i$'s share variance denoted $\sigsq_z$.

It is straightforward to demonstrate that the strategy's mass function $\prob{W}$ is ancillary to the likelihood of the responses and thus the randomization procedure can be ignored during estimation \citep[page 98]{Rosenberger2016}. 


Table~\ref{tab:pop_model_estimators} provides the expression of the two estimators we consider ($\betaThatDiffMeans$ and $\betaThatLinRegr$), their expectations, variances and mean squared errors (MSE). The expectations are taken over the distribution of $\Z$, as this is the source of randomness in the response and thus the expressions are conditional on $\w$ and $\z$. To simplify the expressions we denote the imbalance in $x$ as $B_x := \xbar_T - \xbar_C$ and the imbalance in $z$ as $B_z := \zbar_T - \zbar_C$ where the bar notation denotes sample average and the subscript indicates arm and $r := n^{-1} \sum_{i=1}^n x_i z_i$, a covariance-like term measuring the relatedness of the observed and unobserved measurements. For further simplicity we let $\beta_x = 1$ which simplifies our expressions by removing constants that merely signify the signal ratio of observed to unobserved covariates.

\begin{table}[ht]
\centering
\caption{Estimator properties in the population model.}
\begin{tabular}{c|c|c}
& $\betaThatDiffMeans$ & $\betaThatLinRegr$ \\
\hline
Finite Estimate 	& 	$\betaT + B_x + B_z	$  				& $\betaT + B_x + (B_z - r B_x)(1- B_x^2)^{-1}$  \\
Expectation		& 	$\betaT + B_x$							& $\betaT$ \\
Variance 			& 	$\overn{\sigsq_z}$ 					& $\overn{\sigsq_z} (1- B^2_x)^{-1}$ \\ \hline
MSE 					& 	$ \overn{\sigsq_z} + B_x^2$ 	& $\overn{\sigsq_z} (1- B^2_x)^{-1} $
\end{tabular}
\label{tab:pop_model_estimators}
\end{table}

If the optimal design is selected by the MSE, then the best assignment is the one that minimizes $B_x$, the imbalance in the observed covariate. This is the main justification used to advocate against randomization and instead use deterministic optimal designs \citep{Kiefer1959, Harville1975, Bertsimas2015, Kallus2018}.  This debate date back to the inception of experimentation with \citet{Smith1918} and a good review of classic works including \citet{Kiefer1959} is given by \citet[Chpater 3]{Steinberg1984}. 

Here, the strategy $W$ is degenerate where $\prob{W = \w_*} = 1$ for $\w_*$, the MSE-optimal assignment. To find $\w_*$ in practice is more difficult. One can formulate the procedure as a binary integer programming problem. If forced balance (a term defined in Section~\ref{subsec:restricted_designs}) is required, the partitioning problem is NP hard, meaning that there does not exist a known algorithm that can find the optimal allocation in polynomial time. There are approximations that run in polynomial time that are usually close enough for practical purposes, e.g. branch and bound \citep{Land1960}.

Tangentially, we also observe that if the model is linear, then $\betaThatLinRegr$ is more efficient than $\betaThatDiffMeans$ as its variance is $\approx \sigsq_z /n + (\sigsq_z /n) B_x^2$ when using two terms in a geometric series approximation\footnote{And thus is only appropriate when $\sigsq_z /n \leq 1$.} versus the variance of the simple estimator $\sigsq_z /n + B_x^2$. This will become important when we discuss our methodology in sec.~\ref{sec:methodology}

Once again, those that assume the population model prefer optimal deterministic design.

\subsection{The Randomization Model Perspective}\label{app:randomization_model}

In the randomization model \citep[Chapter 6.3]{Rosenberger2016} also called the \qu{Fisher model} or \qu{Neyman model}, the source of the randomness is in the treatment assignments $\w$. \qu{The $n$ subjects are the population of interest; they are not assumed to be randomly drawn from a superpopulation} \citep[page 297]{Lin2013}. Table~\ref{tab:rand_model_estimators} provides the expression of the estimator, its expectation, variance and mean squared error (MSE) for both the differences-in-means and linear regression estimators assuming only that $\w$ comes from a forced balance procedure. The expectations are taken over the distribution of $W$, as this is the source of randomness in the response and thus the expressions are conditional on $\z$ and $\x$.

\begin{table}[ht]
\centering
\caption{Estimator properties in the randomization model.}
\begin{tabular}{c|c|c}
& $\betaThatDiffMeans$ & $\betaThatLinRegr$ \\
\hline
Finite Estimate 			& 	[same as Table~\ref{tab:pop_model_estimators}]  & [same as Table~\ref{tab:pop_model_estimators}]  \\
Expectation		& 	$\betaT$	& $\betaT + O(1/\sqrt{n})\sqrt{E(B_x^4))}+O(E(B_x^3))$ \\
Variance and MSE 			& $\oneover{n^2} (\x + \z)^\top \bSigmaw (\x + \z)$ & $\left(2\expesub{\w}{B^2_x B^2_z} +  \oneover{n^2} \z_\perp^\top \bSigmaw \z_\perp\right)(1+O(E(B_x^2)))$
\end{tabular}
\label{tab:rand_model_estimators}
\end{table}

The response model of Equation~\ref{eq:simple_model} under the randomization model perspective was studied by \citet{Kapelner2019} and the expressions in Table~\ref{tab:rand_model_estimators} can be found therein. To derive these expressions, there was one additional assumption placed on $W$ that we will make precise in Section~\ref{sec:methodology}: for any assignment $\w$, the assignment where the treatment group subjects are swapped for the control groups subjects, $-\w$, is equally likely. Also note that for $\betaThatLinRegr$, no closed form expressions are available so they are approximated to the third order in a geometric series.

The $\bSigmaw$ term denotes $\var{W}$, the variance-covariance matrix of the strategy that features ones along the diagonal and off-diagonals that gauge the covariance between subject $i$'s assignment and subject $j$'s assignment. $\z_\perp$ is defined to be the component of $\z$ orthogonal to $\x$. It makes sense that this term is important in the error of the linear regression estimator as the linear regression can only adjust for the component of $\z$ it has access to via $\x$.

The most striking observation is that the optimal design strategy is not clear from the MSE expressions as they were in the population model perspective. Our response model was deliberately picked to be the most simplistic and our estimators were chosen to be the most popular. Further, if optimal design were to be defined by minimal MSE, it cannot be resolved as $\z$ is unknown, a practical problem addressed in Section~\ref{sec:methodology}.

We note that there is finite sample bias in the linear regression estimator because the regression assumed a response model that omitted the additive $\z$ term. However, this bias is small as noted by \citet{Freedman2008} and \citet{Lin2013}. 

\subsubsection{The Effect of Large Imbalances in the Observed Covariates}\label{app:unlucky_assignments}

Running the experiment under a particular unlucky assignment is destructive to both perspectives. This can be seen in our simple model of Equation~\ref{eq:simple_model} where the estimator features an additive $B_x$ term. In the population model perspective, the MSE of $\betaThatDiffMeans$ suffers an additive penalty of $B^2_x$ and the MSE of $\betaThatLinRegr$ suffers a multiplicative penalty of $(1 - B^2_x)^{-1}$ derived from the specific unlucky assignment (see Table~\ref{tab:pop_model_estimators}). In the randomization model perspective, the MSE of $\betaThatDiffMeans$ suffers an additive penalty of $n^{-2}\expesub{\w}{B^2_x}$ and the approximate MSE of $\betaThatLinRegr$ also suffers an additive penalty but it is more difficult to see as it is buried in the quartic form $\expesub{\w}{B^2_x B^2_z}$ (see Table~\ref{tab:rand_model_estimators}). These penalties are due to the presence of many unlucky assignment\emph{s} in $\allocspace$.

\subsection{The MSE for $\betaThatDiffMeans$}\label{app:mse_multivariate_simple}

This section is heavily adapted from \citet[Appendices 6.2--6.3]{Kapelner2019}.

We first show that the estimator is unbiased i.e. $\cexpesubnostr{\w}{\betaThatDiffMeans}{\z,\X; \bbeta} = \betaT$. Throughout this appendix $\X$ is considered fixed. Using the model given by Equation~\ref{eq:realistic_response_model}:

\beqn
\cexpesubnostr{\w}{\betaThatDiffMeans}{\z, \X; \bbeta}  &=& \oneover{n} \cexpesubnostr{\w}{\w^\top \parens{\betaT \w + \X\bbeta + \z}}{\z, \X; \bbeta} \\
&=& \oneover{n} \parens{
\expesubnostr{\w}{\betaT \w^\top \w} +
\cexpesubnostr{\w}{\w^\top \X\bbeta}{\X; \bbeta} +
\cexpesubnostr{\w}{\w^\top \z}{\z}
} \\
&=& \oneover{n} \parens{
\betaT \expesubnostr{\w}{\w^\top \w} +
\cexpesubnostr{\w}{\w^\top \X\bbeta}{\X; \bbeta} +
\expesubnostr{\w}{\w}^\top \z
} \\
&=& \betaT + \oneover{n} \parens{
\cexpesubnostr{\w}{\w^\top \X\bbeta}{\X; \bbeta} +
\expesubnostr{\w}{\w}^\top \z
}
\eeqn

\noindent Since $w_i \in \{-1,+1\}$ then, ${\w^\top\w}=\sum_{i=1}^n w_i^2 =n$. The assumption that $W$ is a mirror strategy (Assumption~\ref{ass:mirror}) implies that $\expesubnostr{\w}{\w} = \zerovec_n$ and thus

\bneqn\label{eq:mirror_used_to_show_vanish}
\cexpesub{\w}{\w^\top \X\bbeta}{\X, \bbeta}= 0
\eneqn

\noindent since each $\w$ cancels out with the summand with $-\w$ with shared probability.

This unbiasedness implies that the MSE equals the variance,

\beqn
\cvarsubnostr{\w}{\betaThat}{\z, \X; \bbeta} &=& \cexpesubnostr{\w}{\betaThatDiffMeanssq}{\z, \X; \bbeta} - \beta_T^2 \\
&=& \cexpenostr{(\w^\top \y /n)^2}{\z, \X; \bbeta} - \beta_T^2 \\
&=& \oneover{n^2} \expe{
\squared{
\w^\top (\beta_T \w + \X\bbeta + \z)
}
\,\Big|\, \z, \X; \bbeta} - \beta_T^2\\
&=& \oneover{n^2} \expe{
\squared{
\beta_T \w^\top \w + \w^\top(\X\bbeta + \z)
}
\,\Big|\, \z, \X; \bbeta} - \beta_T^2 \\
&=& \oneover{n^2} \expe{
2n\beta_T \w^\top (\X\bbeta + \z) +
(\w^\top(\X\bbeta + \z))^2
\,\Big|\, \z, \X; \bbeta}
\eeqn

\noindent where the last line follows from $\w^\top \w= n$, simplification and canceling out the constant $\beta_T^2$. By the same arguments of Equation~\ref{eq:mirror_used_to_show_vanish}, $\cexpesub{\w}{\w^\top (\x + \z)}{\z, \X; \bbeta} = 0$ leaving us with

\beqn
&=& \oneover{n^2} \expe{
(\w^\top(\X\bbeta + \z))^2
\,\Big|\, \z, \X; \bbeta} \\
&=& \oneover{n^2} \expe{
(\X\bbeta + \z)^\top \w \w^\top (\X\bbeta + \z)
\,\Big|\, \z, \X; \bbeta} \\
&=& 
\oneover{n^2} 
(\X\bbeta + \z)^\top \bSigmaw (\X\bbeta + \z).
\eeqn

\subsection{The Expectation of the MSE for $\betaThatDiffMeans$}\label{app:expe_mse_multivariate_simple}

This section is heavily adapted from \citet[Appendix 6.4]{Kapelner2019}. We compute

\beqn
\expesubnostr{\z}{\cmsesubnostr{\w}{\betaThat}{\z, \X; \bbeta}} &=& \cexpesub{\z}{\oneover{n^2}(\X\bbeta + \z)^\top \bSigmaw (\X\bbeta + \z)}{\x} \\
&=& \oneover{n^2}  \cexpesub{\z}{(\X\bbeta)^\top \bSigmaw \X\bbeta + 2 (\X\bbeta)^\top \bSigmaw \z + \z^\top \bSigmaw \z}{\X; \bbeta} \\
&=& \oneover{n^2} \bbeta^\top\Xt \bSigmaw \X\bbeta + \frac{2}{n^2} \bbeta^\top\Xt \bSigmaw \cexpesub{\z}{\z}{\X} + \oneover{n^2}\expesub{\z}{\z^\top \bSigmaw \z} \\
\eeqn 

\noindent and note that since $\cexpesubnostr{\z}{\z}{\X} = \zerovec_n$ (by implication of Assumption~\ref{ass:random_z}), the second term is zero. 

The third term is the expectation of a quadratic form. By \citet[Equation 318]{matrixcookbook} and assuming homoskedasticity (Assumption~\ref{ass:homo}) and the fact that since $W$ is generalized multivariate bernoulli, this implies $\tr{\bSigmaw} = n$ we arrive at

\beqn
\expesubnostr{\z}{\cmsesubnostr{\w}{\betaThat}{\z,\x}} = \overn{\sigsq_z} + \oneover{n^2}\bbeta^\top \X^\top \bSigmaw \X \bbeta.
\eeqn 

%
%
%
%

\subsection{The MSE for $\betaThatLinRegr$}\label{app:mse_multivariate}

We let the overall design matrix of the OLS regression be $\Xtilde := \bracks{\w ~|~\X}$. By standard OLS theory and Equation~\ref{eq:realistic_response_model},

\beqn
\bv{\betaThat} = \XtXinvXttilde \y = \twobytwomat{n}{\w^\top \X^\top}{\X\w}{\X^\top \X}^{-1} \twovec{\w^\top \y}{\X^\top \y}
\eeqn

\noindent Note that the first entry of the estimator above is $\betaThatLinRegr$, what we care about. We use formula given in \citet[Section 3.2.6]{matrixcookbook} to provide the top row of the inverse matrix above:

\beqn
\betaThatLinRegr = \parens{\twobytwomat{\oneover{n - \w^\top \P \w}}{\frac{-\w^\top \X (\X^\top \X)^{-1}}{n - \w^\top \P \w}}{\text{N/A}}{\text{N/A}} \twovec{\w^\top \y}{\X^\top \y}}_1 &=& \frac{\w^\top \y - \w^\top \P \y}{n - \w^\top \P \w} \\
&=& \frac{\w^\top (\I - \P) \y}{n - \w^\top \P \w}
\eeqn

\noindent where $\P := \X (\X^\top \X)^{-1} \X^\top$, i.e. the projection matrix onto the covariates and $\I$ is the $n \times n$ identity matrix. Now, replacing $\y$ with the model, we find:

\beqn
\betaThatLinRegr &=& \frac{\w^\top (\I - \P) (\betaT \w + \bbeta \X + \z)}{n - \w^\top \P \w} \\
&=& \frac{\w^\top (\betaT \w + \bbeta \X + \z) - \w^\top \P (\betaT \w + \bbeta \X + \z) }{n - \w^\top \P \w} \\
&=& \frac{n\betaT + \w^\top \bbeta \X + \w^\top \z - \betaT \w^\top \P \w - \w^\top \bbeta \X - \w^\top \P \z}{n - \w^\top \P \w} \\
&=& \frac{\betaT  (n - \w^\top \P \w) + \w^\top \z - \w^\top \P \z}{n - \w^\top \P \w} \\
&=& \betaT  + \frac{\w^\top (\I - \P) \z}{n - \w^\top \P \w} \\
&=& \betaT  + \frac{B_z - \B_{\X}^\top \A  \r}{1 - \B_{\X}^\top \A  \B_{\X}} \\
&=& \betaT  + \underbrace{\frac{B_z -   \BXprime \r'}{1 - \BXprimet \BXprime}}_g \\
\eeqn

\noindent where $\r = \X^\top \z / n$, a column vector of correlation-like metrics of the $p$ covariates with $\z$, denote $\B_{\X} = \X^\top \w / n$, a column vector of average differences in the $p$ covariates between treatment and control, $\A = n(\X^\top \X)^{-1}$, a symmetric matrix, $\BXprime := \A^{\half} \B_{\X}$, $\r' := \A^{\half} \r$ and $B_z$ is the same as in the univariate case, the average difference in $z$ between treatment and control. Note that $\BXprimet \BXprime$ is the Mahalanobis distance between the average covariate values in the treatment group and the average covariate values in the control group.

We can show that $g$ simplifies to $\frac{B_z - r B_x}{1 - B_x^2}$ in the univariate case (the expression in Table~\ref{tab:rand_model_estimators}). Note that this estimator is not a function of $\bbeta$! This is a convenient result of covariate adjustment.

The $\cmsesubnostr{\w}{\betaThatLinRegr}{\z}$ is the expectation of $(\betaThat - \betaT)^2 = g^2$. This expectation is difficult to express in closed form. We instead take a geometric series approximation,


\beqn
g = \frac{B_z -   \BXprime \r'}{1 - \BXprimet \BXprime} &=& \parens{B_z - \BXprime  \r'} \parens{1 + \normsq{\BXprime} + \Oof{\normfourth{\BXprime}}}
\eeqn

\noindent We can use asymptotic notation to expand $g^2$ and then approximate it

\beqn
g^2 &=& \squared{B_z - \BXprime  \r'} \squared{1 + \normsq{\BXprime} + \Oof{\normfourth{\BXprime}}} \\
&=& \parens{\squared{B_z - \BXprime  \r'} + 2 \normsq{\BXprime} B_z^2} \parens{1 + \Oof{\normsq{\BXprime}}} \\
&\approx& \squared{B_z - \BXprime  \r'} + 2 \normsq{\BXprime} B_z^2
\eeqn

\noindent Recall that our target is $\cmsesubnostr{\w}{\betaThat}{\z} = \expesubnostr{\w}{g^2}$. The expectation of the first term can be written as

\beqn
\expesub{\w}{\squared{B_z - \BXprime  \r'}} &=& \expesub{\w}{\squared{\w^\top (\I - \P) \z}} \\
&=& \expesub{\w}{(\z (\I - \P))^\top \w \w^\top (\I - \P) \z} \\
&=& \z^\top (\I - \P) \bSigmaw (\I - \P) \z \\
&=& \z^\top \G \z
\eeqn

\noindent where $\G := (\I - \P) \bSigmaw (\I - \P)$. This term can alternatively be expressed as $\z^\top_\perp \bSigmaw \z_\perp$ where $\z_\perp$ is defined to be the component of $\z$ orthogonal to the column space of $\X$.

\noindent The expectation of the second term is:

\beqn
\expesub{\w}{2 \normsq{\BXprime} B_z^2} &=& 2 \expesub{\w}{\normsq{\BXprime} B_z^2} \\
&=& 2 \expesub{\w}{B_z \oneover{n} \w^\top \P \w B_z} \\
&=& \frac{2}{n^3} \expesub{\w}{\z^\top \w \w^\top \P \w \w^\top \z} \\
&=& \frac{2}{n^3} \z^\top \D \z
\eeqn

\noindent where $\D := \expesub{\w}{\w \w^\top \P \w \w^\top }$, an expectation of a homogeneous quartic form that cannot be simplified further. Putting this all together,

\beqn
\cmsesubnostr{\w}{\betaThatLinRegr}{\z} &=& \oneover{n^2}  \z^\top \G \z + \frac{2}{n^3} \z^\top \D \z \\
&=& \z^\top \R \z
\eeqn

\noindent where the determining matrix of the quadratic form is $\R := \G + \frac{2}{n}\D$.

%
%
%
%
%
%
%
%

\subsection{The Expectation of the MSE for $\betaThatLinRegr$}\label{app:expe_mse_multivariate}

We find the expectation over $\z$ as

\beqn
\expesub{\z}{\cmsesub{\w}{\betaThatLinRegr}{\z}} = \oneover{n^2} \expesub{\z}{\z^\top \R \z} =  \frac{\sigsq_z}{n^2} \tr{\R}
\eeqn

\noindent where the equality comes from Assumptions~\ref{ass:independence} and \ref{ass:homo} and an application of \citet[Equation 318]{matrixcookbook}.

\beqn
\tr{\R} &=& \tr{\G} + \frac{2}{n} \tr{\D} \\
&=& \tr{(\I - \P) \bSigmaw (\I - \P)} + \frac{2}{n} \tr{\expesub{\w}{\w \w^\top \P \w\w^\top}} \\
&=& \tr{(\I - \P) \bSigmaw} + \frac{2}{n} \expesub{\w}{\tr{\w \w^\top \P \w\w^\top}} \\
&=& n - \tr{\P\bSigmaw} + \frac{2}{n} \expesub{\w}{\tr{\w\w^\top\w \w^\top \P }} \\
&=& n - \tr{\P\bSigmaw} + 2 \expesub{\w}{\tr{\w \w^\top \P }} \\
&=& n - \tr{\P\bSigmaw} + 2\expesub{\w}{\w^\top \P \w} \\
&=& n - \tr{\P\bSigmaw} + 2 \tr{\P\bSigmaw}  \\
&=& n + \tr{\P\bSigmaw}\\
&=& n + \tr{\XXtXinvXt \bSigmaw} \\
&=& n + \tr{\XtXinvXt \bSigmaw \X} \\
&=& n + \tr{\XtXminussqrt \Xt \bSigmaw \X \XtXminussqrt} \\
&=& n + \oneover{n} \tr{A^{\half} \Xt \bSigmaw \X A^{\half} } \\
&=& n + \tr{\X_\perp^\top \bSigmaw \X_\perp }.
\eeqn

Let $\X_\perp := \X \XtXminussqrt$ be the orthogonalization of $\X$. Thus,

\beqn
\expesub{\z}{\cmsesub{\w}{\betaThatLinRegr}{\z}} &=& \frac{\sigsq_z}{n} + \frac{\sigsq_z}{n^2} \tr{\X_\perp^\top \bSigmaw \X_\perp }
\eeqn

\noindent where the trace term can more intuitively be expressed as:

\beqn
\tr{\X_\perp^\top \bSigmaw \X_\perp } &=& \x_{\perp_{\cdot 1}}^\top \bSigmaw \x_{\perp_{\cdot 1}} + \ldots + \x_{\perp_{\cdot p}}^\top \bSigmaw \x_{\perp_{\cdot p}} \\
&=& \expesub{\w}{B^2_{\x_{\perp_{\cdot 1}}}} + \ldots + \expesub{\w}{B^2_{\x_{\perp_{\cdot p}}}}
\eeqn

\noindent meaning the sum of imbalances squared of the each of the orthogonalized dimensions of the column space of $\X$.

\subsection{The Standard Error of the MSE for $\betaThatDiffMeans$}\label{app:se_mse_multivariate_simple}

This section is heavily adapted from \citet[Appendix 6.6]{Kapelner2019}.

Section~\ref{app:mse_multivariate_simple} derived the quadratic form of the MSE. The variance with respect to $\z$ can be calculated straightforwardly using \citet[eq. 319]{matrixcookbook} when assuming that the fourth moment of $\z$ is finite (Assumption~\ref{ass:finite_fourth}) and do not depend on $\X$ (Assumption~\ref{ass:high_moments_indep}) as

\bneqn\label{eq:mse_as_function_of_z}
\varsub{\z}{\cmsesubnostr{\w}{\betaThat}{\z,\x}} = \frac{\sigsqzsq}{n^4} 
\parens{
	n \kappa_z + 2 \frobsq{\bSigmaw} + \frac{4}{\sigsq_z} \bbeta^\top \Xt \bSigmaw^2 \X\bbeta +\gamma_z \onevec_n^\top \bSigmaw \X \bbeta
}, \nonumber
\eneqn

\noindent where $\kappa_z$ is the excess kurtosis in $Z$ and $\gamma_z := \expe{z^3}$ and we prove in sec.~\ref{deriv:skewness_term_zero} that this last term $\gamma_z \onevec_n^\top \bSigmaw \X\bbeta$ is zero (regardless of skew) if we assume~\ref{ass:forced_balance} that all assignments are forced balance.\\

\begin{assumption}[Forced Balance] \label{ass:forced_balance} For all $w \in \allocspace$, $\w^\top \onevec_n = 0$.
\end{assumption}

Since the rerandomization strategy in the text employed BCRD as the base strategy, this assumption was implicit throughout the paper.

\subsection{A Proof that the Last MSE Term is Zero}\label{deriv:skewness_term_zero}

This section is heavily adapted from \citet[Appendix 6.7]{Kapelner2019}. We wish to demonstrate that $\onevec_n^\top \bSigmaw \X\bbeta = \bbeta^\top \X^\top \bSigmaw \onevec_n = 0$. By the forced balance assumption (\ref{ass:forced_balance}), $\cexpesubnostr{\w}{\onevec_n^\top \w}{\X,\bbeta} = \varnostrsub{\w}{\onevec_n^\top \w\,|\,\X,\bbeta} = 0$ since every $\w$ is balanced. Then, $\varnostrsub{\w}{\onevec_n^\top \w\,|\,\X,\bbeta} = \cexpesubnostr{\w}{\squared{\onevec_n^\top \w}}{\X,\bbeta} = \onevec_n^\top \bSigmaw \onevec_n = 0$.

Note that $\bSigmaw = \sumonen{i}{\lambda_i \v_i \v_i^\top}$ where the $\lambda_1 \geq \ldots \geq \lambda_n \geq 0$ and $\v_i$'s are its eigenvalues and eigenvectors respectively. Since $\bSigmaw$ is a variance-covariance matrix, it is symmetric implying that its eigenvalues are all non-negative. We can then write $\onevec_n^\top \bSigmaw \onevec_n = \sumonen{i}{\lambda_i \squared{\onevec_n^\top \v_i}} = 0$. This means that $\lambda_i \squared{\v_i^\top \onevec_n} = 0$ for all $i$. In order for this to be true, for every $i$ either $\lambda_i = 0$ or $\v_i^\top \onevec_n = 0$.

We now examine just the term $\bSigmaw \onevec_n$ which can be written as $\sumonen{i}{\lambda_i \v_i \v_i^\top \onevec_n}$. For all $i$ either $\lambda_i$ or $\v_i^\top \onevec_n$ is zero rendering the \qu{middle} $\v_i$ irrelevant. Thus $\bSigmaw \onevec_n = \zerovec_n$ and $\bbeta^\top \X^\top \bSigmaw \onevec_n = \bbeta^\top \X^\top \zerovec_n = 0$.

\subsection{The Standard Error of the MSE for $\betaThatLinRegr$}\label{app:se_mse_multivariate}

By Assumptions~\ref{ass:independence} and \ref{ass:homo} and an application of \citet[Equation 319]{matrixcookbook},

\beqn
\varsub{\z}{\oneover{n^2} \z^\top \R \z} = \frac{\sigsqzsq}{n^4} \parens{2 \tr{\R^2} + \kappa_z \sum_{i=1}^n R_{i,i}^2}
\eeqn

\noindent where $\kappa_z$ denotes excess kurtosis and $R_{i,i}$ are the diagonal entries of $\R$ that we cannot simplify further. We evaluate the trace term below:

\small
\beqn
\tr{\R^2} &=& \frobsq{\G + \frac{2}{n}\D} \\
&=& \tr{ \parens{\G + \frac{2}{n}\D} \parens{\G + \frac{2}{n}\D} } \\
&=& \tr{\G^2 + \frac{4}{n}\G\D + \frac{4}{n^2}\D^2} \\
&=& \tr{\G^2} + \frac{4}{n}\tr{\G\D} + \frac{4}{n^2}\tr{\D^2} \\
&=& \tr{(\I - \P) \bSigmaw (\I - \P)(\I - \P) \bSigmaw (\I - \P)} + \frac{4}{n}\tr{(\I - \P) \bSigmaw (\I - \P)\D} + \frac{4}{n^2}\tr{\D^2} \\
&=& \tr{(\I - \P) \bSigmaw (\I - \P) \bSigmaw} + \frac{4}{n}\tr{(\I - \P) \bSigmaw (\I - \P)\D} + \frac{4}{n^2}\tr{\D^2} \\
&=& \frobsq{(\I - \P) \bSigmaw} + \frac{4}{n}\tr{\G\D} + \frac{4}{n^2}\tr{\D^2}. \\
\eeqn
\normalsize

\noindent Putting this all together, we find that

\small
\beqn
&&\sesub{\z}{\cmsesub{\w}{\betaThat}{\z, \X}} = \frac{\sigsq_z}{n^2} \sqrt{2 \parens{\frobsq{(\I - \P) \bSigmaw} + \frac{4}{n}\tr{\G\D} + \frac{4}{n^2}\tr{\D^2}} + \kappa_z \sum_{i=1}^n R_{i,i}^2}.
\eeqn
\normalsize

We now prove some bounds on some of these terms so we can interpret them more easily:

\beqn
\tr{\D^2} \leq \tr{\D}^2 = n^2 \tr{\X_\perp^\top \bSigmaw \X_\perp }^2
\eeqn

\noindent where the first inequality is a trace inequality for symmetric matrices and the second equality was shown in App.~\ref{app:expe_mse_multivariate}. Also,

\beqn
\tr{\G\D} \leq \sqrt{\tr{\G^2}\tr{\D^2}} \leq \tr{\D} \sqrt{\tr{\G^2}} = n \tr{\X_\perp^\top \bSigmaw \X_\perp} \frob{(\I - \P) \bSigmaw} 
\eeqn

\noindent where the first inequality is Cauchy-Schwartz, the second is the trace inequality used above and the equality is then by substitutions.

\subsection{Estimating the Squared Frobenius Norm without Bias}\label{app:unbiased_frobsq}

Consider a $p$-dimensional r.v. $W$ where $\expe{W} = \zerovec_n$. Its variance-covariance $n \times n$ matrix $\bSigmaw = \expe{W W^\top}$. We wish to estimate $\theta := \frobsq{\bSigmaw}$ given $\iid$ samples $\w_1, \ldots, \w_S$. The unbiased sample variance-covariance matrix estimator is:

\beqn
\bSigmawhat = \oneover{S} \sum_{s=1}^S \w_s \w_s^\top
\eeqn

\noindent If we use the naive plug-in estimator for $\theta$ i.e. $\thetahat_{naive} := \frobsq{\bSigmawhat}$ then by Jensen's inequality,

\beqn
\expe{\thetahat_{naive}} > \theta
\eeqn

\noindent unless $W$ is degenerate. We now calculate the bias term below. We denote let $w_{s,j}$ be the $j$th entry in $\w_s$. We note the following identity:

\beqn
\squared{\sum_{s=1}^n w_{s,j} w_{s,k}} = \sum_{s=1}^S \squared{w_{s,j} w_{s,k}} + \sum_{s \neq s'} w_{s,j} w_{s,k} w_{s',j} w_{s',k}
\eeqn

\noindent Taking the expectation of both sides we find:

\beqn
\expe{\squared{\sum_{s=1}^S w_{s,j} w_{s,k}}} &=& \sum_{s=1}^S \expe{\squared{w_{s,j} w_{s,k}}} + \sum_{s \neq s'} \expe{w_{s,j} w_{s,k} w_{s',j} w_{is,k}} \\
\expe{\squared{\sum_{s=1}^S w_{s,j} w_{s,k}}} &=& S \expe{\squared{w_{s,j} w_{s,k}}} + S (S - 1) \underbrace{\squared{\expe{w_{i,j} w_{i,k}}}} \\
\eeqn

\noindent The underbraced term is the contribution to the expected Frobenius norm squared of the $j,k$th entry of $\bSigmaw$ which we call $F_{j,k}$. Isolating this, we obtain:

\beqn
F_{j,k} := \squared{\expe{w_{s,j} w_{s,k}}} = \frac{\expe{\squared{\sum_{s=1}^S w_{s,j} w_{s,k}}} - S \expe{\squared{w_{i,j} w_{i,k}}}}{S (S - 1)}
\eeqn

\noindent Now using plugin estimators we define:

\beqn
\hat{F}_{j,k} := \frac{\squared{\sum_{s=1}^S w_{s,j} w_{s,k}} - \sum_{s=1}^S \squared{w_{s,j} w_{s,k}}}{S (S - 1)}
\eeqn

\noindent which is an unbiased estimator for $F_{j,k}$ since each component is an unbiased estimator. Now, to get the total Frobenius norm estimator which we call $\thetahat$, we sum over all entries:

\beqn
\thetahat = \sum_{j=1}^n \sum_{k=1}^n \hat{F}_{j,k} = \oneover{S (S - 1)} \parens{\sum_{j=1}^n \sum_{k=1}^n \squared{\sum_{s=1}^S w_{s,j} w_{s,k}} - \sum_{j=1}^n \sum_{k=1}^n \sum_{s=1}^S \squared{w_{s,j} w_{s,k}}}
\eeqn

\noindent Note that $\sum_{s=1}^S w_{s,j} w_{s,k}$ is the $j,k$th entry of $S \bSigmawhat$ which means the double sum over its squared is just the Frobenius norm squared of $S \bSigmawhat$, i.e. 

\beqn
\thetahat = \oneover{S (S - 1)} \parens{S^2 \thetahat_{naive} - \sum_{j=1}^n \sum_{k=1}^n \sum_{s=1}^S \squared{w_{s,j} w_{s,k}}}
\eeqn

\noindent This is as far as we can go for general r.v. $W$. Specifically for the case where $W$ is an allocation strategy whose support is $\braces{-1,+1}^n$, then $\squared{w_{s,j} w_{s,k}} = 1$ always. Thus we have:

\beqn
\thetahat = \oneover{S (S - 1)} \parens{S^2 \thetahat_{naive} - n^2 S} = \frac{S}{S-1} \thetahat_{naive} - \frac{n^2}{S-1}
\eeqn

\subsection{Estimating the Squared Frobenius Norm without Bias with a constant multiple}\label{app:unbiased_frobsq_const}

Now we wish to do a different problem. We wish to estimate $\theta := \frobsq{(I-P)\bSigmaw}$ given $\iid$ samples $\w_1, \ldots, \w_n$. Note the following identity

\beqn
\frobsq{\bSigmaw} = \frobsq{(\I-\P)\bSigmaw} + \frobsq{\P\bSigmaw} + 2 \angbrace{(\I-\P)\bSigmaw, \P\bSigmaw}_F
\eeqn

\noindent where 

\beqn
\angbrace{(\I-\P)\bSigmaw, \P\bSigmaw}_F  = \tr{((\I-\P)\bSigmaw)^\top \P\bSigmaw} =  \tr{\bSigmaw^\top(\I-\P) \P\bSigmaw} = 0
\eeqn

\noindent so that

\beqn
\frobsq{(\I-\P)\bSigmaw}  = \frobsq{\bSigmaw} - \frobsq{\P\bSigmaw}
\eeqn

\noindent and we have an unbiased expression for the first term on the right hand side. Now we work on the second term:

\beqn
\frobsq{\P\bSigmaw} = \frobsq{\XXtXinvXt\bSigmaw} = \frobsq{\expe{\XXtXinvXt\w\w^\top}}
\eeqn

\noindent which means the frobenius norm squared's $j,k$ entry contribution will be:

\beqn
F_{j,k} = \squared{\expe{\x_j \XtXinvXt \w \w_k}}
\eeqn

\noindent Letting $\A := \XtXinvXt$, this can be estimated naively (i.e. biased) via

\beqn
\hat{F}_{j,k} = \squared{\frac{1}{S} \sum_{s = 1}^S {\x_j \A \w_s w_{s,k}}}
\eeqn

\noindent To develop a bias correction, we start with the following identity:

\beqn
\squared{\sum_{s = 1}^S {\x_j \A \w_s w_{s,k}}} = \sum_{s = 1}^S \squared{{\x_j \A \w_s w_{s,k}}} + \sum_{s \neq s'} (\x_j \A \w_s w_{s,k}) (\x_j \A \w_{s'} w_{s',k})
\eeqn

\noindent Taking an expectation

\beqn
\expe{\squared{\sum_{s = 1}^S {\x_j \A \w_s w_{s,k}}}} = \expe{ \sum_{s = 1}^S \squared{{\x_j \A \w_s w_{s,k}}} } + \expe{\sum_{s \neq s'} (\x_j \A \w_s w_{s,k}) (\x_j \A \w_{s'} w_{s',k})}
\eeqn

\noindent We then have:

\beqn
\expe{ S^2 \hat{F}_{j,k} } = S \expe{ \frac{1}{S} \sum_{s = 1}^S \squared{{\x_j \A \w_s w_{s,k}}}}  + S(S-1) F_{j,k}
\eeqn

\noindent Re-arranging, this gives us the unbiased estimator:

\beqn
\hat{F}^{*}_{j,k} = \frac{S}{S-1} \hat{F}_{j,k}  -  \frac{1}{S(S-1)} \sum_{s = 1}^S \squared{{\x_j \A \w_s w_{s,k}}} 
\eeqn

\noindent Summing this over $j$ and $k$, thus an unbiased estimator for the whole frobenius norm of the second term is

\beqn
\hat{ \frobsq{\P\bSigmaw} } = \frac{S}{S-1} \frobsq{\hat{ \P\bSigmaw} } - \frac{1}{S(S-1)}  \sum_{j=1}^n \sum_{k=1}^n \sum_{s=1}^S \squared{{\x_j \A \w_s w_{s,k}}} 
\eeqn

\noindent We notice that $w_{s,k}^2 = 1$, and hence can collect terms over k:
 
\beqn
\hat{ \frobsq{\P\bSigmaw} } = \frac{S}{S-1} \frobsq{\hat{ \P\bSigmaw} } - \frac{n}{S(S-1)}  \sum_{j=1}^n \sum_{s=1}^S \squared{{\x_j \A \w_s }} 
\eeqn

\noindent The rightmost term can also be re-written as:

\beqn
\frac{n}{S(S-1)}  \sum_{j=1}^n \sum_{s=1}^S \squared{{\x_j \A \w_s }}& =&  \frac{n}{S-1} \sum_{j=1}^n  \P_j \left[ \frac{1}{S} \sum_{s=1}^S \w_s \w_s^T \right] \P_j^T = \frac{n}{S-1} \sum_{j=1}^n  \P_j \bSigmawhat \P_j ^T\\
&=& \frac{n}{S-1} \tr{  \P \bSigmawhat \P}=\frac{n}{S-1} \tr{  \P \bSigmawhat},
\eeqn
where $\P_j$ is the $j$'th row of $\P$. We apply the bias correction only for $S \ge 30$; otherwise, the variance of the estimator is too high for the bias correction to be meaningful. 

\section{Algorithms}\label{app:algs}
Below are the algorithms for finding $a_*$ discussed in the main body of the manuscript.
\begin{algorithm}[htp]
\caption{This algorithm returns the optimal rerandomization design $\allocspace_*$ along with the optimal threshold $a_*$ and the value of the relative tail criterion for the minimum design $Q'_*$. The argument \qu{TailMSE} takes one of the following functions: TailDistLR for the procedure providing an explicit distribution of Section~\ref{subsec:est_Q_provide}, TailNormalLR for the procedure assuming normality of Section~\ref{subsec:est_Q_normality} and TailApproxLR for the procedure assuming the approximation of Section~\ref{subsec:est_Q_approx} (these functions are found in Algorithm~\ref{alg:lr_tail_functions}). The \qu{...} means other parameters that will be different for each of the tail functions. This procedure is implemented to be faster via caching intermediate values.}
\begin{algorithmic}
\Procedure{OptimalRerandomizationDesign}{$\X$, $\imbal$, $\allocspacebase$, $q$, TailMSE, ...}
	\State $n \gets $nrow($\X$)
	\State $Bs \gets \braces{\imbal(\X,\w)\,:\,\w \in \allocspacebase}$ 		\Comment{Precompute all balances}
	\State $\braces{Bs^{sort}, \allocspacebase^{sort}} \gets$  sort\_asc($Bs$, $\allocspacebase$) \Comment{Sort all balances from smallest to largest and use these sorted indices to sort $\allocspacebase$}
	\State $s_f \gets \text{size}(\allocspacebase)$, $s_* \gets$ NULL, $Q'_* \gets \infty$ \Comment{Initialize search parameters}
	\For {$s = 1 \ldots s_f$} 
		\State $\allocspace_s \gets \allocspacebase^{sort}[1 : s]$ 	\Comment{The brackets indicate the sub-array operator} 
		\State $\bSigmaw \gets \oneover{s} \sum_{\w \in \allocspace_s} \w \w^\top$
		\State $Q' \gets$ TailMSE($\allocspace_s$, $s$, $\X$, $n$, $\bSigmaw$, $q$, ...)
		\If{$Q' <  Q'_*$}
			\State $Q'_* \gets Q'$
			\State $s_* \gets s$
		\EndIf
	\EndFor \\
	\Return new HashMap($\allocspace_* = \allocspacebase^{sort}[s_* : s_f]$, $a_* = Bs^{sort}[s_*]$, $Q'_* = Q'_*$)
\EndProcedure
\end{algorithmic}
\label{alg:master}
\end{algorithm}

\begin{algorithm}[htp]
\caption{This algorithm provides the methods for computing the tail criterion for $\betaThatLinRegr$.}
\begin{algorithmic}
\Function{TailDistLR}{$\allocspace_s$, $s$, $\X$, $n$, $\bSigmaw$, $q$, simZn, $N_Z$} \Comment{Strategy of Section~\ref{subsec:est_Q_provide}}
	\State shared\_lr
	\State $Q's \gets $ new\_array($N_Z$) \Comment{Empty Array of size}
	\For {$n_Z = 1 \ldots N_Z$}
		\State $\z \gets $simZn($n$) \Comment{A user-provided function}
		\State $Q's[n_Z] = \z^\top \parens{\G + \overn{2} \D} \z$
	\EndFor \\
	\Return empirical\_quantile($Q's$, $q$)
\EndFunction
~\\
\Function{TailNormalLR}{$\allocspace_s$, $s$, $\X$, $n$, $\bSigmaw$, $q$} \Comment{Strategy of Section~\ref{subsec:est_Q_normality}}
	\State shared\_lr
	\State $\lambda$s $\gets$ eigen\_values\_of$\parens{\G + \overn{2} \D}$ \\
	\Return Hall\_Buckley\_Eagleson\_Quantile\_Approx($\lambda$s, $q$)
\EndFunction
~\\
\Function{TailMSEApproxLR}{$\allocspace_s$, $s$, $\X$, $n$, $\bSigmaw$, $q$, $\kappa_z$} \Comment{Strategy of Section~\ref{subsec:est_Q_approx}}
	\State shared\_lr
	\State $\X_\perp \gets \X \XtXminussqrt$
	\State $c \gets $ normal\_quantile($q$) \\
	\Return Equation~\ref{eq:tail_criterion_prime_linear_estimator} computed
\EndFunction\\
~\\
shared\_lr$\{$
~~\State $\P \gets \XXtXinvXt$
~~\State $\G \gets (\I - \P) \bSigmaw (\I - \P)$
~~\State $\D \gets \oneover{s} \sum_{\w \in \allocspace_s} \w \w^\top \P \w \w^\top$\\
$\}$
\end{algorithmic}
\label{alg:lr_tail_functions}
\end{algorithm}

\end{document}